\documentclass[acmtog,nonacm]{acmart}
\usepackage[utf8]{inputenc}
\usepackage{booktabs} 

\setcopyright{none}

\usepackage[ruled,noalgohanging,norelsize]{algorithm2e} 

\SetAlFnt{\normalsize}
\SetAlCapFnt{\small}
\SetAlCapNameFnt{\small}
\SetAlCapHSkip{0pt}

\usepackage{amsmath}
\usepackage{amsfonts}
\usepackage{enumitem}

\usepackage{microtype}
\usepackage{units}
\usepackage{soul}
\usepackage{multicol}

\makeatletter
\newcommand{\manuallabel}[2]{\def\@currentlabel{#2}\label{#1}}
\makeatother
\usepackage{graphicx}
\usepackage{caption}
\usepackage{algorithm2e}[vlined, ruled]
\SetKw{AlgName}{Name:}
\SetAlFnt{\ttfamily}
\SetProgSty{texttt}

\SetCommentSty{mycommfont}

\SetAlCapNameFnt{\normalsize}
\SetAlCapFnt{\normalsize}
\DeclareCaptionType{alg}[Algorithm]
\DeclareCaptionLabelFormat{alg}{#1 ̃#2}

\usepackage{multirow}

\usepackage{array}   
\newcolumntype{L}{>{$}l<{$}} 

\usepackage{qtree}

\usepackage[finalizecache=false,frozencache=true,cachedir=.,outputdir=.]{minted}
\newcommand{\mi}[1]{\mintinline{cpp}{#1}}

\usepackage{tikz,pgfplots,pgfplotstable}

\pgfplotsset{
  /pgfplots/ybar legend/.style={
      /pgfplots/legend image code/.code={%
          \draw[##1,/tikz/.cd,yshift=-0.25em]
          (0cm,0cm) rectangle (3pt,0.8em);},
    },
}

\makeatletter
\newcommand\resetstackedplots{
  \makeatletter
  \pgfplots@stacked@isfirstplottrue
  \makeatother
  \addplot [forget plot,draw=none] coordinates{(1,10) (2,10) (3,10)};
}
\makeatother

\usepackage{xcolor}
\definecolor{bblue}{HTML}{4F81BD}
\definecolor{rred}{HTML}{C0504D}
\definecolor{ggreen}{HTML}{9BBB59}
\definecolor{ppurple}{HTML}{9F4C7C}
\definecolor{bbrown}{HTML}{ca7034}

\usepackage[customcolors]{hf-tikz}

\tikzset{style green/.style={
      set fill color=green!50!lime!60,
      set border color=white,
    },
  style cyan/.style={
      set fill color=cyan!90!blue!60,
      set border color=white,
    },
  style orange/.style={
      set fill color=orange!80!red!60,
      set border color=orange!80!red!60,
    },
  hor/.style={
      above left offset={-0.1,0.31},
      below right offset={0.1,-0.125},
      #1
    },
  ver/.style={
      above left offset={-0.27,0.3},
      below right offset={0.1,-0.1},
      #1
    }
}

\usepackage{dsfont}
\usepackage{wrapfig}

\definecolor{myorange}{rgb}{0.87451, 0.521569, 0.262745}
\definecolor{myblue}{rgb}{0.6, 0.6, 1.}
\definecolor{mygreen}{rgb}{0.498039, 0.662745, 0.596078}

\usepackage{mdframed}
\mdfsetup{skipabove=\topskip,skipbelow=\topskip}
\newmdenv[	linewidth=0.75pt,
  leftline=false,
  rightline=false,
  innerleftmargin=0pt,
  innerrightmargin=0pt]{topbot}

\newcommand{\cc}{c}
\newcommand{\disable}[1]{}

\newcommand{\figref}[1]{Figure\ \ref{#1}}
\newcommand{\secref}[1]{Section\ \ref{#1}}

\newcommand{\R}{\mathds{R}}

\newcommand{\device}[1]{{\ttfamily\mdseries\upshape #1}}

\definecolor{darkred}{rgb}{0.7, 0.02, 0.02}
\definecolor{atomictangerine}{rgb}{1.0, 0.6, 0.4}
\definecolor{babyblueeyes}{rgb}{0.63, 0.79, 0.95}
\definecolor{asparagus}{rgb}{0.53, 0.9, 0.42}
\definecolor{lightgrey}{rgb}{0.4, 0.4, 0.4}

\definecolor{revColor}{rgb}{0.12, 0.4, 0.0}

\newcommand{\rev}[1]{#1}

\newcommand{\revv}[1]{#1}

\newcommand{\revvst}[1]{}

\newcommand{\hide}[1]{}
\newcommand{\alter}[1]{}
\newcommand{\opt}[1]{}
\newcommand{\rem}[1]{}

\acmBooktitle{book}

\begin{document}
\title{Sparsity-Specific Code Optimization using Expression Trees}

\author{Philipp Herholz}
\email{ph.herholz@gmail.com}
\affiliation{%
  \institution{ETH Zurich}
  \country{Switzerland}
}

\author{Xuan Tang}
\email{xt585@nyu.edu}
\affiliation{%
  \institution{New York University}
  \country{USA}
}

\author{Teseo Schneider}
\email{teseo@uvic.ca}
\affiliation{%
  \institution{University of Victoria}
  \country{Canada}
}

\author{Shoaib Kamil}
\email{kamil@adobe.com}
\affiliation{%
  \institution{Adobe Research}
  \country{USA}
}

\author{Daniele Panozzo}
\email{panozzo@nyu.edu}
\affiliation{%
  \institution{New York University}
  \country{USA}
}

\author{Olga Sorkine-Hornung}
\email{sorkine@inf.ethz.ch}
\affiliation{%
  \institution{ETH Zurich}
  \country{Switzerland}
}

\setlength{\algomargin}{0pt}

\begin{abstract}
    We introduce a code generator that converts unoptimized C++ code operating on sparse data into vectorized and parallel CPU or GPU kernels. Our approach unrolls the computation into a massive expression graph, performs redundant expression elimination, grouping, and then generates an architecture-specific kernel to solve the same problem, assuming that the sparsity pattern is fixed, which is a common scenario in many applications in computer graphics and scientific computing. We show that our approach scales to large problems and can achieve speedups of two orders of magnitude on CPUs and three orders of magnitude on GPUs, compared to a set of manually optimized CPU baselines.
To demonstrate the practical applicability of our approach, we employ it to optimize popular algorithms with applications to physical simulation and interactive mesh deformation.
\end{abstract}

\maketitle

\section{Introduction} \label{sec:intro}

The optimization of numerical code using combinations of algorithmic improvements and custom hardware played a core role in advancing computer graphics, enabling the move from the early days of rasterizing individual polygons in seconds to the modern era of raytracing entire scenes at interactive rates.  With manufacturers building more specialized processors and increased parallelism in hardware, the path to optimizing code for a specific processor is becoming more complex.

In this work we focus on optimizing numerical code that is dominated by sparse operations. Algorithms in computer graphics, and geometry processing in particular, make heavy use of sparse matrices or can be easily formulated as operations on them. A reason for this fact is that we often study how global behavior of a system emerges from local interactions, which are inherently sparse. This observation generalizes to adjacent disciplines like finite element analysis in engineering and computer vision.

Arithmetic operations between sparse matrices can be quite costly and hard to optimize. In contrast to their dense counterparts, they cannot easily benefit from highly optimized BLAS or LAPACK dense matrix routines. Recent implementations \cite{mkl,cusparse,rocmsparse} of sparse matrix operations, like the sparse matrix-matrix product, enable analyzing the sparsity structure of specific operands and the result in order to schedule an optimized parallel computation plan and preallocate memory. Executing this plan computes these operations efficiently as long as the sparsity structure does not change. Unfortunately, these optimized implementations are very limited. Nvidia's cuSparse \cite{cusparse}, for example, only supports products of sparse matrix in row-major format and does not allow implicitly transposing one of the factors. Sparse matrix addition is not supported directly; only accumulation of sparse matrix products of the same structure is possible. Even more restricting is the fact that compound operations are generally not supported. For example, the computation pattern for the operation $\mathbf L \mathbf M \mathbf L^\top + \mathbf A$, where all operators are sparse matrices, is also fixed but must be broken down into individual operations for the pre-analysis approach.

We take the idea of pre-analysis a step further and analyze full sparse expressions. We assume that we have a set of (sparse) input matrices that are combined to form an output matrix. Our method is not limited to a fixed set of arithmetic operations but allows for arbitrary manipulations and storage formats as long as they are implemented in C++ code. We build on ideas recently introduced with the EGGS system \cite{eggs}. Like EGGS, we represent all values of the output matrix as \emph{expression trees} encoding all operations necessary to compute the output values from the input values.
We decompose, group, and optimize these expression trees to generate code that efficiently evaluates all trees. Our generated programs utilize expression unrolling, vectorization, parallelization, and regular memory accesses, both on CPU and GPU, to achieve 1--3 orders of magnitude in performance increase. 

We extend EGGS in several ways. Instead of operating on entire output expression trees individually, we optimize complex sparse operations by operating on subtrees of finer granularity. This allows us to optimize a spectrum of applications where EGGS would fail to produce faster code. Additionally, we target GPUs and introduce several new optimization steps, including optimizing for coalesced memory accesses and aligned memory reads.

We additionally support sparse matrix construction. Assembling sparse matrices from individual entries can take a significant amount of time, sometimes even dominating linear system solves in finite element based simulations. In the context of physical simulation, for example, the current state of a simulation is characterized by position and velocity data. Our method can generate code that takes these values as input and constructs the final sparse system matrix in one step. In many cases, the sparse matrix is the Hessian of a non-linear energy, for example when employing Newton's method. Our system supports constructing the Hessian directly from an energy formulation using symbolic automatic differentiation. We show that code generated by automatic differentiation is as efficient as code generated from a hand optimized implementation of the Hessian.

Storing runtime information in the form of expression trees allows us to generate fine tuned code; however, this approach comes with some disadvantages. A central limitation is the focus on static expressions which means that we have only limited support for branching based on non-constant numerical values. Moreover, the memory requirements for storing all expressions can be significant. We report peak memory usage of our code generator in Table \ref{tab:timings} and demonstrate that our method is still able to scale to typical mesh resolutions. In terms of preprocessing time our system requires up to 30 minutes in extreme cases which makes it useful only if many instances of the code will be executed, for example in a commercial application. We believe that an optimized and parallelized implementation of our pipeline as well as improved expression caching can alleviate these problems. We elaborate on application areas in \secref{sec:requirements}.

We validate our approach by automatically optimizing the performance of popular computer graphics algorithms, e.g. for mesh deformation \cite{arap} and physical simulation \cite{Baraff:1994}, by using our approach on existing code. This automatic optimization only requires minor modifications to the original code in order to apply our approach.

\subsection{Contributions}
In this paper we present several algorithmic and technical contributions that improve upon EGGS \cite{eggs}. The most important novel features are: 
\setlength\itemsep{0.5em}%
\begin{enumerate}[leftmargin=0em]
    \item[] \textbf{Expression decomposition.} We introduce advanced expression decomposition (\secref{sec:decomposition}). By breaking down expression trees into smaller expressions that can be cached and reused, we address a central problem of EGGS: the lack of scalability in terms of expression complexity.

    \item[] \textbf{Memory optimization.} We introduce several ways to optimize memory accesses (\secref{sec:backend}) and demonstrate their practical utility, especially when targeting GPUs. 

    \item[] \textbf{Code optimization.}
    We present new ways to optimize generated code using expression simplification (\secref{sec:simplification}) and code transformation to facilitate vectorization (\secref{sec:backend}). Moreover, we support  code generation for evaluating derivatives.
\end{enumerate}

In an ablation study (\secref{sec:ablation}) we demonstrate that all contributions cooperatively result in the significant speedups we report. Depending on the example and target device, the effects of individual optimizations vary.

We will publish the code of our system to facilitate further research. We consider our prototype implementation to be a proof of concept demonstrating that it is possible to accelerate a wide range of example programs even with limited scalability in terms of  processing time, memory and code characteristics. We discuss those limitations (e.g. Section~\ref{sec:requirements}, Appendix~\ref{sec:codeTimings}) and plan to continue development towards software applicable in even broader scenarios.

\section{Related Work}
\paragraph{Dense \& sparse matrix libraries}
Libraries for dense linear algebra~\cite{LAPACK,atlas,eigen,numpy,armadillo,mkl}
have found wide adoption due to their ability to optimize computation using
platform-specific code.  One of the most popular recent packages is the 
BLIS~\cite{BLIS1} library generator, which utilizes
the combination of data movement specification and a single kernel definition
to generate optimized code for dense matrix multiplication, using the approach by \citet{goto}.

Sparse matrices are increasingly important for applications, often written in languages
such as MATLAB~\shortcite{matlab}, Julia~\cite{julia},
and with libraries like Eigen~\cite{eigen}, PETSc~\cite{petsc}, Blaze~\cite{iglberger2012}, SciPy~\cite{scipy}
and Boost uBLAS~\cite{ublas}.
Different libraries support different sparse matrix formats for different
subsets of operations, with compressed sparse row/column (CSR/CSC) and
coordinate (COO) being most common.  For some applications, leveraging
blocked structure within the sparse matrix can lead to better performance,
as first demonstrated by OSKI and the parallel pOSKI libraries~\cite{oski, byun2012},
which use the block CSR (BCSR) format to reduce memory traffic and better utilize
registers.
More recently, sparse neural networks~\cite{Liu_2015_CVPR} have made libraries
for sparse computations important in machine learning. Tensorflow Sparse~\cite{tensorflow-sparse}  
and TorchSparse~\cite{tang2020searching} introduce optimized sparse computations
to existing machine learning frameworks.

\paragraph{Sparse linear algebra compilers}
Compilers targeting sparse linear algebra generate customized code based on
the linear algebra operation and data structure.  Bik and Wijshoff~\shortcite{dutch1,
  dutch2} and the Bernoulli project~\cite{bernoulli} built compilers for certain sparse
operations.  More recently, \citet{venkat2016} utilize the CHiLL polyhedral
framework~\cite{chill} to optimize sparse matrix-vector operations using an inspector-executor
approach and the polyhedral model~\cite{FEAUTRIER1988ARRAY,PUGH1991UNIFORM} to tailor runtime behavior to the specific matrix. The polyhedral model allows for compact representation and optimization of a program without resorting to explicit loop unrolling.

The taco tensor algebra~\cite{taco} compiler generalized previous approaches to sparse code generation
to build a compiler that supports a wide variety of tensor formats~\cite{taco-stephen} for
generating mixed sparse and dense tensor algebra.  The compiler has been extended to support
Halide-style~\cite{halide-siggraph, halide-pldi} separation of computation from scheduling, enabling taco to generate optimized code for both CPUs and GPUs~\cite{taco-ryan}.  The techniques we present could
be used to extend taco to support code generation tailored for specific sparsity patterns.

\paragraph{Sparsity-specific code generation}
Prior work has applied polyhedral techniques to generate efficient sparsity-specific code
by finding dense substructures within sparse matrices~\cite{augustinepldi2019} for SpMV.
In contrast, our work extends beyond sparse matrix-vector multiplication.
Sparsity-specific code generation techniques have also been used to speed up Cholesky factorization
and backsolves~\cite{cheshmi}; the technique utilized for these solves analyzes dependencies between values in order to create an execution plan, while our technique relies on per-entry expression trees.  Finally, while we build upon the CPU-specific code generation demonstrated
by EGGS~\cite{eggs}, our technique extends and generalizes their approach to support GPUs
and includes new optimizations to better utilize memory bandwidth and generate more efficient code.
RXMesh~\cite{rxmesh} uses a domain-specific strategy to partition meshes into contiguous patches, allowing
GPUs to better utilize coalesced memory accesses.  Our strategy optimizes memory access patterns by grouping expressions.

\paragraph{\revv{Domain specific languages for simulation \& geometry processing}}
\revv{In recent years several domain specific languages (DSLs) for phyiscal simulation and geometry processing have been introduced. Their development is commonly motivated by the idea to separate the implementation of on algorithm operating on a (mesh) graph and its \emph{scheduling} on a specific device, similar to Halide \cite{halide-siggraph} for image processing. 
Simit \cite{simit} and Ebb \cite{ebb} provide languages that make it convenient to express global energies in terms of local contributions. This approach enables the user to focus on the actual energy formulation and automatically generate correct, efficient code for CPUs or GPUs. The Opt system \cite{opt} and its recent successor Thallo \cite{thallo} follow a similar route but focus on the class of non-linear least squares problems. These methods can automatically evaluate the required Jacobians and generate complete solvers based on the Gauss-Newton or Levenberg-Marquardt algorithm. These method have a different focus compared to our approach. They strive to make the implementation of specific energies more convenient and focus on the optimization of solvers and very specific sparse operations. 
The power of these domain specific languages stems from their high level of abstraction making them convenient to use, at the cost of limited generality. Our approach, on the other hand, is agnostic to the fact that there may be an underlying mesh structure, and is potentially complementary to DSLs for specific uses of sparse matrices. The other key difference is that we do not try to generate efficient code for a specific type of optimization problem but rather for a specific problem instance.}

\section{Overview}\label{sec:method0}
\begin{figure*}
    \includegraphics[width=0.9\linewidth]{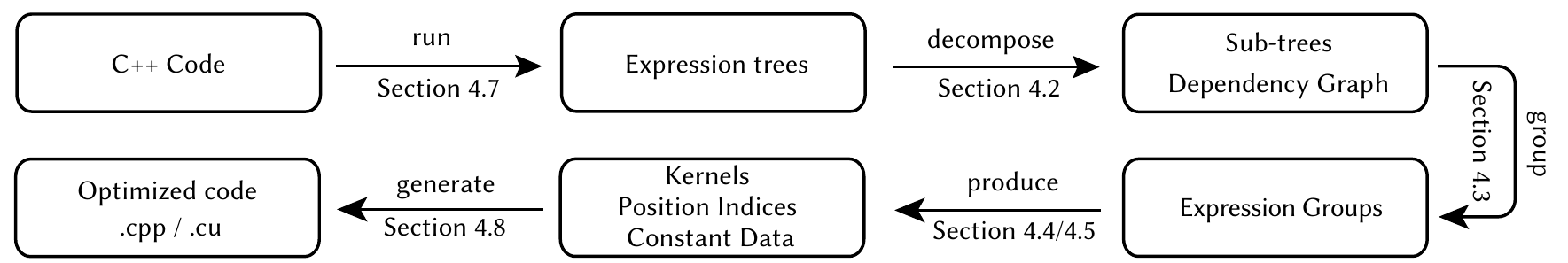}
    \caption{Overview of our code optimization technique. Starting from an existing implementation we generate code that is optimized for a specific structure of input values. We target both, CPUs and GPUs.}
    \label{fig:overview}
\end{figure*}

\label{sec:overview}
\setlength{\columnsep}{4pt}

In this section we first provide background information about concepts introduced in EGGS \cite{eggs} and used by our system. We then give an overview of the different stages of our method and elaborate on alternative implementation options. Finally we describe the scope of applications that benefit from our approach the most.

\subsection{Background}
We take C++ code containing algebraic operations on sparse matrices and their construction as input for our method. The operations can be implemented using Eigen~\cite{eigen} or any other templated library in combination with custom code. We build upon the basic principles employed in the EGGS system
and rely on the concept of \textit{expression trees} to encode numerical computations.
To facilitate the integration of such a system with existing algorithms, both EGGS' and our approach replace numerical types (e.g., \texttt{double}) with a custom \texttt{Symbolic} class, which uses operator overloading
to effectively build an expression tree for every output value, deferring the computations to a later stage.
As an example, consider the following code implementing a simple function \texttt{f}:
\begin{minted}{cpp}
    Symbolic g(Symbolic a, Symbolic b) {
        return a + b;
    }

    Symbolic f(Symbolic a, Symbolic b) {
        return 2 * g(a, b) + sqrt(a);
    }
\end{minted}

\noindent Executing it with \texttt{Symbolic} arguments representing the variables \texttt{a} 
and \texttt{b} will return the expression tree depicted in the inset. If the 
\begin{wrapfigure}[8]{l}{0.19\linewidth}
    \vspace{-12pt}
    \includegraphics[width=\linewidth]{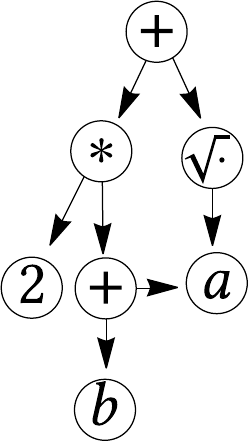}
\end{wrapfigure}
function \texttt{f} is part of an existing non templated code base,
a minor change of the code is required: template \texttt{f} and \texttt{g} to support execution with floating point as well as with
\texttt{Symbolic} arguments. Each instance is either a constant, variable or an operation on a set of child instances. All instances together with their parent/child relationships define a \emph{directed acyclic graph} (DAG), a data structure commonly used in compiler design \cite{aho2013}. We differentiate between the concept of expression tree and its efficient encoding as a DAG. 
Executing the function \mi{h}, for example, will result in the DAG depicted to the right.

\begin{minipage}[c]{0.7\linewidth}
\begin{minted}[tabsize=2]{cpp}
Symbolic h(Symbolic a, Symbolic b) {       
    x = a * (a + b);
    return x * x;              
}
\end{minted}
\end{minipage}\hfill
\begin{minipage}[c]{0.29\linewidth}
    \hspace{22pt}%
	\includegraphics[width=0.5\linewidth]{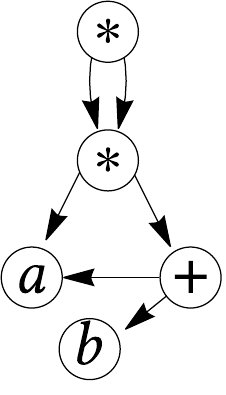}%
\end{minipage}
\par\noindent Note that the leaf \mi{a} and the expression \mi{x} are only stored once making it easy to identify common subexpressions. The full expression tree is not explicitly represented and would contain individual nodes for
these expressions.
While the trees of different output values may differ, they commonly share subtrees that have 
the same structure in the following sense: Two trees are structurally equivalent if 
they only differ 
by their leaves but not by the operations required to evaluate them.
Consequently, we can compute the two results using the same piece of (optimized) code which we call \emph{kernel}.
\begin{wrapfigure}[7]{l}{0.3\linewidth}%
    \vspace{-10pt}%
   	\centering%
	\includegraphics[width=\linewidth]{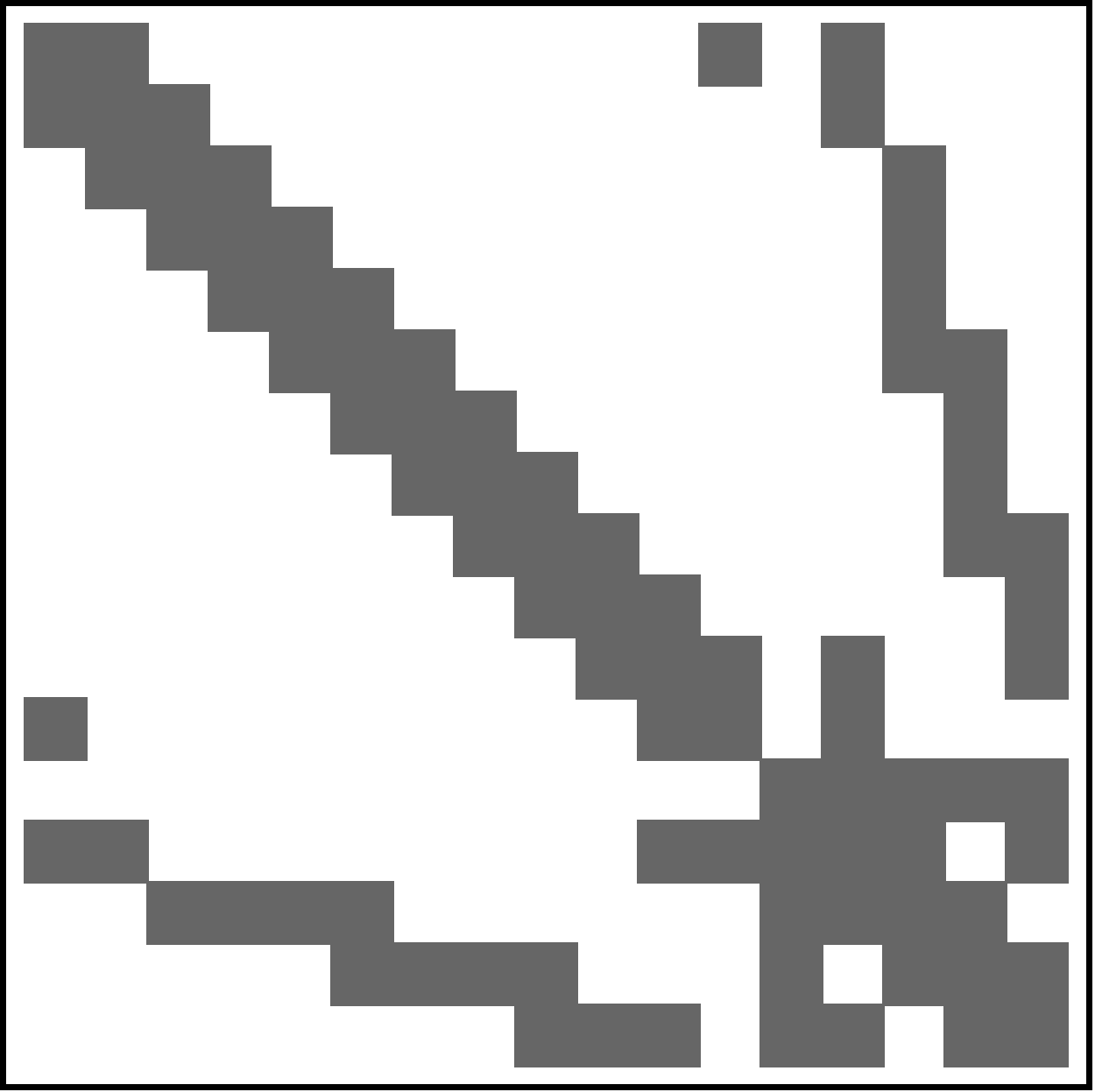}%
\end{wrapfigure} 
A common source for structurally similar expressions are sparse matrix operations. The inset shows the sparsity pattern of the Laplacian $\mathbf L$ of a mesh with 17 vertices. In order to compute the matrix product $\mathbf L^2 = \mathbf L^\top \mathbf L$ efficiently, only the non-zero
values of $\mathbf L^2$ are computed. Computing each element of $\mathbf L^2$ from the entries $l_i$ of $\mathbf L$ means evaluating expressions of the form
\begin{align*}
	l_{p_1}  l_{p_2} + \cdots + l_{p_{n-1}}  l_{p_n},
\end{align*}
where $p_i$ is an index referencing a particular entry of the input matrix. 
Since the matrix inherits its structure from a mesh, the length of these sums differ only within a certain range; in this case $n \in \{6,8,10,12,14,16,18\}$ \rev{takes on only 7 different values}. Our method will generate code for computing the matrix product by running 7 vectorized kernels in parallel; one for each possible value of $n$. For $n = 6$ the code for the kernel will have the following form
\begin{minted}[tabsize=2]{cpp}

x[p[7*i]] = x[p[7*i+1]]*x[p[7*i+2]] 
  + x[p[7*i+3]]*x[p[7*i+4]] + x[p[7*i+5]]*x[p[7*i+6]];

\end{minted}
This example does not feature any subexpressions that appear multiple times \rev{ and can be efficiently handled by EGGS as well. The computation of $\mathbf L^3$, however, is more involved and causes EGGS to fail producing performant code (see \secref{sec:structured}) while our method succeeds.}

\subsection{System overview}
A CPU kernel evaluating several instances of structurally equivalent expressions might look like the following code:
\begin{minted}[tabsize=2]{cpp}
	void k(double*x, double* c, int* p) {
		for (int i = 0; i < 256; ++i) {
		    x[p[2*i]] = c[i] * sqrt(x[p[2*i+1]]);
		}
	}
\end{minted}
Each loop iteration evaluates one of the 256 expression trees making up the group. The array \mi{p} determines where the values of leaf variables are stored and \mi{c} contains the constants.
Kernels can significantly benefit from vectorization and parallelization, as we describe in \secref{sec:backend}. The goal of our method is to evaluate all expressions of a specific program using a small number of kernels.
Since we compile fixed expression trees, our technique is limited in terms of conditional branching if the condition depends on numeric values, as we discuss in \secref{sec:code_execute}.

Transforming a given piece of code together with a set of symbolic input and output values into an optimized program requires several steps (Figure~\ref{fig:overview}). To provide an overview, we briefly describe them here and follow up with more details in the next section.%
\setlength\itemsep{1em}%
\vspace{1em}%
\begin{enumerate}[leftmargin=0em]
    \item[] \textbf{Symbolic execution} (Section~\ref{sec:code_execute}). The first step executes the code to be compiled. After this phase, each output variable contains an expression tree that can be used to produce the output value for an arbitrary assignment of the input variables. Optionally, we can compute symbolic derivatives for specific expressions (Section~\ref{sec:autoDiff}).

    \item[] \textbf{Expression decomposition} (Section~\ref{sec:decomposition}). 
    After the execution phase we end up with a set of expressions. By counting the references to each node and analyzing their complexity we decide which subexpressions should be evaluated independently with their result stored in variables (\emph{intermediate} expressions) 

    \item[] \textbf{Expression grouping} (Section~\ref{sec:grouping}). All outputs and intermediate expressions are grouped based on their structure. We can evaluate all group members using the same \emph{template expression} by substituting the correct leaves. \emph{Kernels} for each group are generated from these expressions.
    In order to facilitate parallelization no group member can depend on a result of another member. We enforce this property by splitting groups if necessary.

    \item[] \textbf{Leaf harvesting} (Section~\ref{sec:leaf-harvesting}). The expressions within a group differ only by their leaves. To identify the leaves, we traverse all group members and store variables and constants in a table with dimensions `number of expressions' $\times$ `number of leaves' for each group. \revvst{With this information we can also identify constants or variables that have the same value for all expressions in the group.}
    
    \item[] \textbf{Expression optimization} (Section~\ref{sec:simplification}). The template expression is hierarchically divided into subexpressions, similar to the expression decomposition stage. Each subexpression will be simplified using a set of algebraic simplification rules.

    \item[] \textbf{Code generation} (\secref{sec:backend}). The last step produces code for a specific target architecture. To this end, we determine the memory locations to store the values of intermediate variables in order to optimize memory access patterns. The code is optimized for vectorization and parallelization.
\end{enumerate}

\subsection{Requirements for input code} \label{sec:requirements}
Apart from being implemented using templates on the numeric type, an algorithm should posses a few characteristics in order to benefit from our approach. 
\setlength\itemsep{1em}
\begin{itemize}[leftmargin=0pt]
    \item[] \textbf{1. Limited conditional branching.} Branching should only depend on constant parameters such as matrix dimensions or should be formulated as a conditional assignment (\secref{sec:code_execute}).
    \item[] \textbf{2. Overhead amortization.} Since the goal of our method is to compile code for a specific structure of input data, we will only see a benefit for applications that run enough instances with the same input structure to amortize the initial compilation cost.
    \item[] \textbf{3. Expression tree variation.} The set of expression trees associated with the output values should feature a small set of unique expression tree structures to facilitate single instruction, multiple data (SIMD) parallelism and a small number of kernels.
\end{itemize}

Several applications in computer graphics, geometry processing, and scientific computing in particular exhibit precisely these properties.
For instance, in the minimization of a non-linear energy defined for a mesh, each step of the iterative procedure requires the computation of the current energy value as well as derivative information, namely the gradient and Hessian. Sufficiently smooth energies will have gradients and Hessians expressed with minimal conditional branching.
Another common property of energies is that they can be computed as the summations of local interactions between mesh elements in applications like bending, shearing, or stretching. Consequently, the Hessian is not only sparse, but the expression trees of their numeric values produce only a limited number of unique tree topologies based on the vertex degree. These properties make a large class of geometric optimization algorithms perfect candidates for our method.
Even if algorithms do not qualify our method might still be applicable. Collision handling in physical simulation, for example, modifies system matrices locally in a way that is unknown at compile time. However, it is in many cases possible to modify a generated system matrix in a preprocess to reflect collision forces. In this situation our method would still optimize the bulk of computation and leave the parts that require conditional branching to the host code.

In \secref{sec:evaluation} we demonstrate that our approach can be applied to a variety of existing codebases. In the next section, we describe the details of our system.

\section{Method}\label{sec:method}
Efficiently handling massive expression trees requires carefully designed algorithms in order to maintain a sufficient degree of scalability. Here we describe the core ideas necessary to implement our method. Throughout we make use of established techiques like directed acyclic graphs and memoization.

\subsection{Efficient tree processing}\label{sec:toolbox}
We need to be able to efficiently compare expression trees in terms of their \emph{structural} and \emph{algebraic} equivalence.
Two trees are \emph{structurally equivalent} (Section~\ref{sec:struct-hashing}) if they are identical up to their leaves, which may contain different symbolic values or constants.
Trees are \emph{algebraically equivalent} (Section~\ref{sec:algHash}) if they evaluate to the same result.
Structurally equivalent trees are algebraically equivalent if their leaves are identical; algebraic equivalent trees do not have to be structurally equivalent. To identify both types of tree equivalence, we leverage a hashing technique that does not require tree traversal. The idea is to compute hashes that only depend on the structure or algebraic equivalence class and identify them uniquely. This can be done by using hash functions generating hashes of sufficient length, like a variant of the secure hash algorithm (SHA), to reduce the collision probability to practically zero. The same concept is used in universally unique identifiers (UUIDs). 
Our implementation allows to optionally check for hash collisions and
we did not find any for 64-bit hashes in our experiments. For other tasks in our pipeline, like dependency detection, tree traversals are not avoidable. In these cases we use \emph{selective} post- or pre-order tree traversals (Section~\ref{sec:traversal}).

\subsubsection{Structural hashing}\label{sec:struct-hashing}
\begin{algorithm}[t!]

\AlgName{structureHash(SymbolicExpression x)}\\
\KwOut{\textrm{Structural hash value for expression x.}}
\vspace{4pt}

\uIf(\tcp*[f]{return hash if it has been computed.}){hashIsValid}{\Return h}
\Else{

\uIf{Op(x) == Variable}{h = variableHash}
\uElseIf{Op(x) == Constant}{h = constantHash}
  \Else{
 \ForEach{c in Childs(x)}
 {h = Hash(structureHash(c), h)}}
   \Return Hash(Hash(Op(x)), h)
 }

\caption{Structure Hashing}
\label{alg:structurehashing}
\end{algorithm}

To compute hashes identifying the structure of an expression we recursively combine subtree hashes. Algorithm~\ref{alg:structurehashing} shows an implementation of this idea. The structural hash value of all leaf nodes is one of two fixed hash values: one for variables and one for constants. All other nodes represent arithmetic operations involving one or more child expressions. The function \texttt{Hash} produces a hash value based on its arguments. 
The structural hash of each expression is computed and stored as soon as the expression is generated.
For commutative operations structural hashes are invariant under child permutation because
we sort them by hash during construction.

\subsubsection{Algebraic hashing}\label{sec:algHash}
Algebraic hashing works similarly to structural hashing but serves a different purpose. If algebraic hashes of two expression trees are identical, they evaluate to the same numeric value. To this end, we only change the way hashes are computed for leaves. For variables, we compute a hash unique to its variable id.
The hash for integer constants is their numerical value. Details and pseudocode can be found in Appendix \ref{app:algHash}.

The hash for expressions representing multiplication or addition are the product or sum of their operands.
The special rules for commutative operations and constants allow us to identify expressions with vastly different tree structures that evaluate to the same value. For instance, the two expressions
$$2  (x + y)  (z + w ) \quad \text{and} \quad 2 x z + (2 z + 2 w) y + x w + w x, $$
will have the same algebraic hash.

\revv{\subsubsection{DAG compression} \label{sec:memoized}
Using hashing we can implement \emph{memoized constructors}. To this end we maintain a map relating hashes to expressions. Whenever a new expression is constructed we first check if an equivalent expression already exists and reuse it if possible.}

\subsubsection{Expression complexity}
\label{sec:complexity}
When optimizing expression trees, we make decisions based on expression complexity $\cc$. 
Whenever we construct an expression, we compute $\cc$ by combining the complexities of its children. For internal nodes, we sum the complexity of their children and add a constant cost based on the node's type.
These values can be adapted based on clock cycles needed for the operations on specific devices.

\subsubsection{Expression traversal}\label{sec:traversal}
Analyzing a set of expression trees cannot solely rely on hashing. We must traverse them to analyze dependencies or to identify the set of leaves for a group of structurally equivalent expressions. In principle, traversing trees can be done in several ways. For example, evaluating a symbolic tree for specific value assignments requires a post-order traversal. Substituting subexpressions by intermediate values, on the other hand, requires a pre-order traversal. In both cases, we proceed in a depth-first manner.
We use \emph{memoization} \cite{aho2013} which amounts to caching the results of visiting each DAG node and reuse them if possible. Without this method most of our processing steps would be infeasible, even for small programs.

\subsection{Expression decomposition}\label{sec:decomposition}
\hide{\begin{figure}
\centering
        \colorbox{gray!7}{
           x : \Tree[.+  [ a  [ a b ].+ ].*  !\qsetw{0.6cm} [.$\exp(\cdot)$ a ] ] 
        } \hspace{6pt}\colorbox{gray!7}{
           y: \Tree[.+  [.$\sqrt{\cdot}$ b ] [ a  [ a b ].+ ].* !\qsetw{0.9cm} ]
        }
        
         \colorbox{gray!7}{
           z : \Tree[.*  [ a  [ a b ].+ ]]
        }
         \colorbox{gray!7}{
           x : \Tree[.+  [ a  [ a b ].+ ].*  !\qsetw{0.6cm} [.$\exp(\cdot)$ a ] ] 
        } \hspace{6pt}\colorbox{gray!7}{
           y: \Tree[.+  [.$\sqrt{\cdot}$ b ] [ a  [ a b ].+ ].* !\qsetw{0.9cm} ]
        }

    \caption{Several trees sharing a common subtree (left). Evaluating this subtree separately and storing the result leads to more efficient programs.}
    \label{fig:treeReference}
\end{figure}}

\begin{figure}
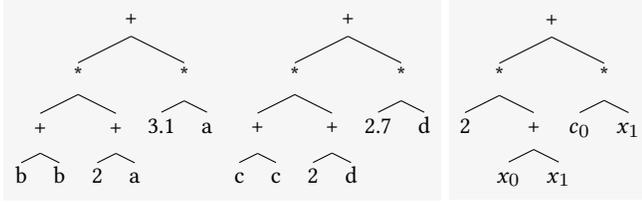

    \begin{minipage}{\linewidth}
        \colorbox{gray!7}{
            \begin{tabular}{cc}
                \hspace{-8pt}
                \Tree[.+  [ [ b b ].+ [ 2 a ].+ ].*  !\qsetw{1.5cm} [ 3.1 a ].* ] & \hspace{-4pt} \Tree[.+  [ [ c c ].+ [ 2 d ].+ ].*  !\qsetw{1.5cm} [ 2.7 d ].* ]
                \hspace{-4pt}
            \end{tabular}}
        \hspace{-1pt}
        \colorbox{gray!7}{
            \hspace{-4pt}
            \begin{tabular}{c}
                \hspace{-7pt}
                \Tree[.+  [ 2 [ $x_0$ $x_1$ ].+ ].*  !\qsetw{1.5cm} [ $c_0$ $x_1$ ].* ]
                \hspace{-4pt}
            \end{tabular}
            \hspace{-4pt}}
    \end{minipage}
    \caption{Two structurally equivalent trees can be evaluated using the optimized tree on the right. Several leafs are identical in each tree and can therefore be encoded by the same variable $x_1$. This enables specific simplifications. Constants can vary per instance ($c_0$) or be constant for all of them ($2$) and can therfore be encoded as a constant floating point value in the kernel template.}
    \label{fig:treeTemplate}
\end{figure}

The constructed output expressions could be directly grouped and compiled into functions.
This can be an effective strategy for some applications, as demonstrated by \citet{eggs}.
Their approach analyzes each group's template expression and stores the result of subtrees that appear multiple times in local variables.
This method is limited in two principal ways. First, it does not decompose redundant subtrees further to detect common subtrees of subtrees --- a situation that is actually quite common in practice. Second, subtrees shared among different output expressions cannot be precomputed. These two major shortcomings lead to higher computation times and result in a potentially excessive number of kernels.

We introduce two levels of expression decomposition. On a \emph{local} level we consider the template expression of each kernel. We identify subtrees that exceed a complexity threshold and appear multiple times in the expression. These subtrees will be evaluated only once with their result stored in a local stack variable for further reference. 
On a \emph{global} level, we consider all output expressions and find subtrees that are shared by more than one of them potentially leading to faster execution times.
Again we perform a hierarchical decomposition. Since these intermediates are used by different kernels and kernel instances, we have to store intermediate results in global memory where they can be reused by any expression referencing the corresponding subtree. 
To implement the two decomposition stages we make use of the DAG formed by the expressions.

\subsubsection{Global decomposition}
\label{sec:globalDecompose}
Intermediate expressions should have a few properties: They should be complex enough and should be referenced several times in order to justify the extra effort of storing and loading their results. To find a `good' partition we adopt a greedy strategy. First we count how often each node is referenced when traversing the DAG starting at the output expressions. All nodes that have been referenced a sufficient number of times $t_\text{ref}$ (we use $t_\text{ref} = 2$ in our experiments) are potential candidates to become intermediate expressions. Many intermediate candidates will be quite small considering that they reference other intermediates. We solve this problem by traversing the DAG again and greedily merge intermediates that are below a complexity threshold $t_\text{compl}$. The DAG now induces a \emph{dependency graph} among all intermediate and output expressions which we can use to identify \emph{local} intermediates. These are nodes that are referenced several times but only within a single output expression. In this situation we do not need to store and load a value from global memory but can evaluate the expression locally. We remove these intermediates and rely on the local decomposition phase to handle them.

\subsubsection{Local decomposition} Each expression belonging to a kernel group can be evaluated using the same \emph{template expression}. We employ local decomposition to identify redundant subexpressions similarly to the global decomposition step. Local intermediates do not have to be stored in global memory and can be directly written to local stack variables. The evaluation order for local intermediates is determined by computing a topological ordering of the dependency graph. In \secref{sec:ablation} we demonstrate that our local decomposition can lead to significant performance improvements and exceeds the capabilities of common compilers. Moreover, it allows for much more compact code.

\subsubsection{Example} We demonstrate both kinds of decomposition using a toy example. We assume that we have three expressions that differ in structure forming three groups, each with one member only. The kernels generated from the respective template expressions are:
\begin{minted}{cpp}
    void k1() {out[0] = sqrt(a*a+b*b)+a;}
    void k2() {
     out[1] = c*(a+d)*sqrt(a*a+b*b)+c*(a+d)+sin(a+d);}
    void k3() {out[2] = a*a+b*b+a*b;}
\end{minted}
Global analysis will detect that the expression \mi{sqrt(a*a+b*b)} occurs two times and we can evaluate it separately as an intermediate value. Moreover, the subexpression \mi{a*a+b*b} appears two times as well, once as part of the first intermediate and once as part of the last output expression. Note that we count it only two times because we only traverse unique subtrees. While traversing the second output expression, we do not traverse the expression \mi{sqrt(a*a+b*b)} because it has been visited before. On a local level, we see that the second expression, stored in \mi{out[1]}, has some redundant subtrees: \mi{c*(a+d)} as well as \mi{a+d} appear multiple times. As a result we have
\begin{minted}{cpp}
    void i1() {r[0] = a*a+b*b;}
    void i2() {r[1] = sqrt(r[0]);}
    void k1() {out[0] = r[1]+a;}
    void k2() {
        double x0 = a+d;
        double x1 = c*x0;
        out[1] = x1*r[1]+x1+sin(x0);
    }
    void k3() {out[2] = r[0]+a*b;}
\end{minted}
For this illustrative example we did not apply a complexity threshold. Introducing global intermediates makes it necessary to invoke the functions in a valid order. In this case \mi{i1()} needs to be called before \mi{i2()} followed by the kernels in arbitrary order.

\subsubsection{Explicit intermediate tagging} \label{sec:explicitIntermediates}
Our automatic decomposition steps consider  expressions individually, however, it might make sense to group several expressions and evaluate them within the same kernel.
A common scenario is the assembly of a sparse stiffness matrix from dense matrices that constitute per-element contributions. The stiffness matrix in a cloth simulation, for example, is assembled from $9\times 9$ blocks for each face. Each entry of this dense block might reuse a quantity (e.g. triangle area) that is not required in any other computation. At runtime the intermediate value representing triangle area has to be loaded 81 times \--- once for each element of the dense matrix.
To solve this issue we offer to manually tag sets of values in the code as \emph{blocks}.
Tagged blocks will be considered as a compound expression and a quantity, like triangle area, will be stored in a local stack variable by the local decomposition stage.
For all our experiments in \secref{sec:evaluation} we explicitly state if and how we made use of tagging. As a toy example consider the following decomposition including one intermediate and two kernels, each with one instance only: 
\begin{minted}{cpp}
    void i1() {r[0] = sqrt(a*a+b*b);}
    void k1() {out[0] = b*r[0]+a;}
    void k2() {out[1] = a*r[0]+b;}
\end{minted}
If the two output variables have been tagged our system generates a single kernel:
\begin{minted}{cpp}
    void k() {
        double y = sqrt(a*a+b*b);
        out[0] = b*y+a;
        out[1] = a*y+b;
    }
\end{minted}
This version avoids the costly loading and storing of a value from global memory.

\subsection{Expression grouping}\label{sec:grouping}
After global expression decomposition, we have a set of output and intermediate expressions.
Our goal is to group them by structural hashes so that each group member can be evaluated using the same \emph{Kernel}. 
Additionally, the group members need to obey the dependency graph (see \secref{sec:decomposition}). To this end we only group expressions if they have the same \emph{level} in the dependency graph. the level of a node in the dependency graph is the length of the shortest path from any output expression to the node.

\subsection{Leaf harvesting}\label{sec:leaf-harvesting}
The group members share the same expression tree structure and only differ by their leaves \figref{fig:treeTemplate}, left).
We \emph{harvest} these leaves by traversing all trees at once, starting at the root and storing all leaves in a table (see Appendix~\ref{app:lhe} for a complete example).
Trees can have millions of nodes. Therefore the implementation of the harvesting step makes benefits from the selective traversal introduced in \secref{sec:traversal}.

\subsection{Expression simplification}\label{sec:simplification}
After identifying the leaves, we can form and optimize a \emph{template expression} that will be used to evaluate the result for all group members. Since we identified some leaves  that are always identical, we have unique optimization opportunities that are not available when compiling a function without any runtime information. Computing the determinant of a matrix, for example, can be heavily optimized if specific elements contain a zero for all group members.
We implemented a set of algebraic simplification transformations, including factoring sums, reducing fractions, summand elimination and evaluation of constant expressions. We describe all of them in Appendix~\ref{app:simplifcationRules}.
Some of the transformations eliminate obvious redundancies, like simplifying a squared norm, that most experienced programmers would try to avoid in the first place. However, optimization opportunities are likely when expressions combine results from independent function calls or for expressions that have been generated by symbolic automatic differentiation.
The ability to simplify these expressions allows us to compete with hand-optimized code (\secref{sec:autodiff-comp}).
Expression transformations are guaranteed to result in arithmetically equivalent expressions; however, this does not mean that they are equivalent when evaluated using floating-point arithmetic. In many cases, this effect will result in an acceptable difference in the order of machine precision; however, it might also introduce (or even prevent) catastrophic numerical cancellation.
To prevent the expression optimizer from introducing numerical cancellation, it is sufficient to disable the optimization of sums.

\subsection{Symbolic type}\label{sec:implementation}

We implemented our method in C++ without any external dependencies. A central challenge is memory management.
Due to the possibly high redundancy of subtrees across all expression trees, we would like to store them only once.
To store repeated expressions only once (DAG), our symbolic numeric type holds a reference counting smart pointer to a hidden datatype storing the actual expression data. Besides storing that reference, the numeric type is responsible for overloading the required operators. This way, using the same expression in the construction of multiple result values uses only one instance of the expression tree. \rev{We provide a basic implementation of our \mi{Symbolic} type in Appendix \ref{app:type}}. We overload a set of basic operations provided by C++ and its standard library. It is straightforward to add additional operations by overloading the corresponding function calls and introducing a unique operation id, name, cost $\cc$ and differentiation rules.
The smart pointer based implementation enables a convenient implementation of memoized constructors (see \secref{sec:memoized}). To this end we optionally maintain a hash map relating algebraic hashes to expressions. When constructing a new expression we reuse equivalent existing expressions if possible. This variant is more costly due to hash map lookups but can provide a significantly lower memory footprint.

\subsection{Symbolic program execution} \label{sec:code_execute}
In order to execute existing code with our symbolic type, all numeric types that directly depend on input data have to be symbolic too. Code that is already templated can directly be used in many cases. This applies to most of Eigen, for example. In all other cases, the numeric types in the code have to be replaced by template parameters (or the \mi{auto} keyword).
Since our method assumes that the code path is independent of the input data, we can only handle limited conditional branching. To this end, conditionals must be replaced by \mi{select(x, a, b)} statements that implement conditional assignment. For numeric types this function template translates to the C++ statement \texttt{x < 0 ? a : b}.
The templated code can be executed with symbolic as well as with numeric types in which case no overhead occurs. This makes the development and debugging of algorithms targeting GPUs and parallel CPU execution very convenient. During algorithm design a function can be tested with basic numeric types, finally calling the function with symbolic types allows to generate optimized code for a desired device.

\subsection{Code generation}\label{sec:backend}
The final step of our pipeline generates the kernel code that will be executed on the target device.
We can now exploit all the runtime information we have about the expression groups to target the target devices' performance features efficiently. We compute all results of a group using two nested loops: an outer one that is executed in parallel and an inner one that facilitates vectorization. Additionally, we can optimize the memory layout of constant data as well as some variables. In the previous steps, we only identified (intermediate) variables and constants and their dependencies but have not decided on a specific position of this data in an array. Our method can use these degrees of freedom to optimize memory access.

To illustrate code generation and variable placement we use a small toy example. We consider the template expression
\begin{minted}{cpp}
  y0 = c0 * x0 * x1 + 2 * c1 * x2 * sqrt(x0 * x1)
\end{minted}
which has been optimized based on 256 group members.

\paragraph{Basic code generation} The symbolic form of the expression can be directly converted into the following kernel code
\begin{minted}[tabsize=4]{cpp}
	for (int i = 0; i < 256; ++i) {
		double x0 = x[p[3 * i]];
		double x1 = x[p[3 * i + 1]];
		double x2 = x[p[3 * i + 2]];
	    double r = x0*x1;
	    x[1024 + i] = c[2*i]*r+2.*c[2*i+1]*x2*sqrt(r);
	}
\end{minted}
The optimizer detected the expression \mi{x0*x1} as redundant and stores it in the temporary variable \mi{r}.
Access to variable data is available through the array \mi{x} while constant data is handled differently. Since constants are unique to group instances, we store them in an array \mi{c} in consecutive order such that they can be easily accessed based on the loop index \mi{i}.
Variable data can only be indexed indirectly since it is shared among all kernels. To this end, we use an index array \mi{p} that stores these addresses. An exception to the indirect addressing is the destination \mi{x[1024 + i]} holding the results. Since the variable positions for intermediate results can be arbitrary as long as they are unique to each variable, we choose them such that they can be written consecutively if this is possible. The array offset \mi{1024} is chosen by the code generator.

\paragraph{Exploiting address coherence} Let us assume that the group we are considering originates from computing a per-vertex quantity based on three coordinates represented by $x_0$, $x_1$, and $x_2$. Chances are that their positions in memory, explicitly stored in \mi{p}, are correlated. They could, for example, be positioned at constant offsets. We analyze if such a property holds for all group members and can simplify the code in the following way
\begin{minted}[tabsize=4]{cpp}
	for (int i = 0; i < 256; ++i) {
		double x0 = x[p[i]];
		double x1 = x[p[i] + 1];
		double x2 = x[p[i] + 2];
	    double r = x0*x1;
	    x[1024 + i] = c[2*i]*r+2.*c[2*i+1]*x2*sqrt(r);
	}
\end{minted}
Consequently, we need to access only $1/3$ of all position indices, which helps limit the use of memory bandwidth at runtime as well as storage requirements.

\paragraph{Coalesced memory access} GPUs benefit tremendously from coalesced memory access. This means that memory access is consecutive across instances. The current memory layout for constants looks like this:
\renewcommand{\arraystretch}{1.2}
\begin{center}
	\vspace{6pt}
	\begin{tabular}{|L|L|L|L|L|L|L|L|L|}\hline
		0     & 1     & 2     & 3     & 4     & 5     & 6     & 7     & \cdots \\\hline
		c_0^0 & c_1^0 & c_0^1 & c_1^1 & c_0^2 & c_1^2 & c_0^3 & c_1^3 & \cdots \\\hline
	\end{tabular}
	\vspace{6pt}
\end{center}
\renewcommand{\arraystretch}{1}
and the command \mi{c[2*i]} accesses the elements at even positions $0$, $2$, $4$ and so on while \mi{c[2*i + 1]} accesses the uneven elements. Since we are free to choose any order we can reorder to allow for coalesced access in the following way
\renewcommand{\arraystretch}{1.2}
\begin{center}
	\vspace{6pt}
	\begin{tabular}{|L|L|L|L|L|L|L|L|L|L|}\hline
		0     & 1     & 2     & 3     & \cdots & 256   & 257   & 258   & 259   & \cdots \\\hline
		c_0^0 & c_0^1 & c_0^2 & c_0^3 & \cdots & c_1^0 & c_1^1 & c_1^2 & c_1^3 & \cdots \\\hline
	\end{tabular}
	\vspace{6pt}
\end{center}
\renewcommand{\arraystretch}{1}
and change the code accordingly
\begin{minted}[tabsize=4]{cpp}
	x[1024+i] = c[i]*r+2.*c[256+i]*x2*sqrt(r);
\end{minted}
Now the constants are accessed consecutively across loop iterations resulting in the desired coalesced memory access. We perform the same reordering for access of the position array \mi{p}. Note that the destination memory is already accessed in a coalesced manner in this example.

\paragraph{CPU vectorization and parallelization}
Finally we restructure the loop such that the C++ compiler can optimally combine vectorization and parallelization. To this end, we use two nested loops. The outer loop runs in parallel, either employing Intel threading building blocks (TBB) or OpenMP pragmas. The inner loop is automatically vectorized if we enable AVX2 auto vectorization, which supports 256 bit registers processing blocks of 4 doubles at the same time. The C++ compiler will not be able to verify that the data we load does not overlap with the memory address in which we store the result. If this were the case, the results would differ between vectorized and non-vectorized code. However, all group members are independent by construction, as detailed in \secref{sec:grouping}, and we can safely tell the compiler to ignore possible dependencies using a pragma statement. For clang the final code after all optimization stages  looks like this:
\begin{minted}[tabsize=2]{cpp}
	tbb::parallel_for(0, 64, [&](int i) {
		#pragma ivdep
		for(int j = 0; j < 4; ++j) {
			double x0 = x[p[4*i+j]];
			double x1 = x[p[4*i+j] + 1];
			double x2 = x[p[4*i+j] + 2];
			double r = x0*x1;
    		x[1024+4*i+j] = \
    			c[4*i+j]*r+2.*c[256+4*i+j]*x2*sqrt(r);
		}
	});
\end{minted}
We initially tried to generate vectorized code using intrinsics directly. However, we found that compilers can generate code that is as efficient or better as long as we provide the correct pragmas. Additionally, relying on the C++ compiler allows our technique to automatically benefit from future or rarely supported vector instruction sets without modifying our code generator.

\paragraph{GPU parallelization}
Since our method is based on the concept of grouping expressions into kernels, it is relatively easy to generate GPU kernels that can be compiled with Cuda or Hip. These compilers automatically generate a program that efficiently exploits vectorization in the form of wavefronts and parallelization, which makes our code generation even easier.

\subsection{Symbolic differentiation}\label{sec:autoDiff}

The technique of automatic differentiation has been used for decades to evaluate differentials of values produced by existing computer programs. 
The symbolic expression trees used in our system lend themselves to be used in conjunction with this established technique. More details on automatic differentiation can be found in the classic textbook of Griewank and Walther~\shortcite{griewank2008evaluating}.

In contrast to operator overloading based forward- and reverse- mode automatic differentiation, we can analyze derivatives in our pipeline and do not have to rely on costly tape data structures at runtime. Since our system systematically caches common subexpressions of derivatives and function values, we automatically address the problem of exponential expression growth. A similar approach is employed in the Opt system \cite{opt}. 
To generate symbolic derivatives, we follow the principle of reverse-mode automatic differentiation since we build derivative expressions by traversing the expression tree from its root. In Appendix \ref{sec:appendixAutoDiff} we provide pseudocode for our implementation.

\section{Evaluation}\label{sec:evaluation}

We evaluate our method on a variety of examples. We compare to alternative methods including manual code optimization, EGGS \cite{eggs}, and dedicated hardware libraries for basic sparse matrix arithmetic. We tested our generated code on different target architectures and compilers to demonstrate its versatility. For all experiments we enable the compiler to perform optimizations \texttt{-O3} and use relaxed floating point rules \texttt{--ffast-math} as well as vectorization \texttt{-mavx2}, which allows us to attribute performance improvements to our method as opposed to the compiler itself.  The following setups have been used

\vspace{2pt}
\begin{center}
\begin{tabular}{ll}
	 \device{Intel} & Intel 2.3 GHz 8-Core Intel Core i9, clang with tbb\\
	 \device{AMD} & AMD EPYC 1.5 GHz 32-Core, gcc with OpenMP\\
	 \device{Vega 20} & AMD Vega 20, 16 GB memory, 3840 ALUs, Hip\\
	 \device{GTX 1080} & GeForce GTX 1080, 8 GB memory, Cuda.
\end{tabular}%
\vspace{2pt}\\%
\end{center}%
Our method performs consistently well on all platform/compiler combinations listed above.
In all cases, we average timings over 100 runs. For the CPU performance, we measure the reference timings on the same machine using the same compiler settings that we use to time the optimized program. In Appendix \secref{sec:codeTimings} we list code generation timings and memory requirements.
We compare against fairly optimized research code. In general it is not easy to define what code is considered to be optimized and in many cases there is potentially the option to employ a even higher degree of manual optimization. For this reason the speedups have to be considered in relation to the original implementation which we provide for our example applications. Our goal is to demonstrate what performance boosts are possible on typical code bases. We hope that users of our open source implementation will report performance boosts for a wider set of applications to get a broader picture of the potential of our method.

\subsection{Sparse matrix operations}
\begin{figure}[t!]
	\begin{center}
		\pgfplotstableread{
size eigen eggs1 	us1 	eggs16 	us16 	gpu	
100k 1 	   13 		35.1 	64.9 	236. 	1139.54
500k 1	   12.93    13.684	70.45 	116.5	1121.
1M	 1 	   13.62	14.29	73.77	117.2	1050.
}\dataNumericA
\pgfplotstableread{
size eigen eggs1 	us1 	eggs16 	us16 	gpu	
100k 1 	   15.28 	23.19	103.16	156.6	1775.5
500k 1	   9.37		19.42	104.7	148.7	2381.8
1M	 1 	   14.86	19.3	114.29	133.21	2665.18
}\dataNumericB
\pgfplotstableread{
size eigen eggs1 	us1 	eggs16 	us16 	gpu	
100k 1 	   12.16	31.89	45.58	344.17	1622.1 
500k 1	   13		19.48	52.		167.	2261
1M	 1 	   14.2		19.3	66.59	151.83	2567.18
}\dataNumericC
\begin{center}
\begin{tikzpicture}
    \begin{axis}[
    	title={$(\alpha \mathbf A + \mathbf B)^\top (\beta \mathbf B^\top + \mathbf C) $},
    	title style={at={(0.5,0.92)},
                   anchor=north,
                   fill=white},
        width  = 1.\linewidth,
        height = 0.5*\linewidth,
        major x tick style = transparent,
        ybar=2pt,
        ymode=log,
        bar width=8pt,
        ymajorgrids = true,
        symbolic x coords={100k,500k,1M},
        xtick=\empty,
        scaled y ticks = false,
        enlarge x limits=0.25,
        enlarge y limits=0.35,
        ymin=0,
        legend cell align=left,
        legend style={
                at={(0.95, 0.75)},
                anchor=north east,
                column sep=1ex
        }
    ]
        \addplot[style={bblue!60,fill=bblue!60,mark=none}] table [meta=size, y=eggs1] {\dataNumericA};
        \addplot[style={rred!60,fill=rred!60,mark=none}] table [meta=size, y=us1] {\dataNumericA};
 		\addplot[style={bblue,fill=bblue,mark=none}] table [meta=size, y=eggs16] {\dataNumericA};
        \addplot[style={rred,fill=rred,mark=none}] table [meta=size, y=us16] {\dataNumericA};
        \addplot[style={ggreen,fill=ggreen,mark=none}] table [meta=size, y=gpu] {\dataNumericA};

    \end{axis}
\end{tikzpicture}
\begin{tikzpicture}
    \begin{axis}[
      	title={$\mathbf A \mathbf B \mathbf C $},
        title style={at={(0.5,0.92)},
                   anchor=north,
                   fill=white},
        width  = 1.\linewidth,
        height = 0.5*\linewidth,
        major x tick style = transparent,
        ybar=2pt,
        ymode=log,
        bar width=8pt,
        ymajorgrids = true,
        symbolic x coords={100k,500k,1M},
        xtick = \empty,
        scaled y ticks = false,
        enlarge x limits=0.25,
        enlarge y limits=0.35,
        ymin=0,
        legend cell align=left,
        legend style={
                at={(0.95, 0.75)},
                anchor=north east,
                column sep=1ex
        }
    ]
        \addplot[style={bblue!60,fill=bblue!60,mark=none}] table [meta=size, y=eggs1] {\dataNumericB};
        \addplot[style={rred!60,fill=rred!60,mark=none}] table [meta=size, y=us1] {\dataNumericB};
 		\addplot[style={bblue,fill=bblue,mark=none}] table [meta=size, y=eggs16] {\dataNumericB};
        \addplot[style={rred,fill=rred,mark=none}] table [meta=size, y=us16] {\dataNumericB};
        \addplot[style={ggreen,fill=ggreen,mark=none}] table [meta=size, y=gpu] {\dataNumericB};

    \end{axis}
\end{tikzpicture}
\begin{tikzpicture}
    \begin{axis}[
      	title={$(\mathbf A + \mathbf B) (\mathbf A \mathbf B + \mathbf C) $},
        title style={at={(0.5,0.92)},
                   anchor=north,
                   fill=white},
        width  = 1.\linewidth,
        height = 0.5*\linewidth,
        major x tick style = transparent,
        ybar=2pt,
        ymode=log,
        bar width=8pt,
        ymajorgrids = true,
        symbolic x coords={100k,500k,1M},
        xtick = data,
        scaled y ticks = false,
        enlarge x limits=0.25,
        enlarge y limits=0.35,
        ymin=6,
        legend columns = 2,
        legend style = {at={(0.5, -0.23)}, anchor=north, inner sep=3pt, style={column sep=0.15cm}},       
   	 	legend cell align=left,
    ]
        \addplot[style={bblue!60,fill=bblue!60}] table [meta=size, y=eggs1] {\dataNumericC};
        \addplot[style={rred!60,fill=rred!60,mark=none}] table [meta=size, y=us1] {\dataNumericC};
 		\addplot[style={bblue,fill=bblue,mark=none}] table [meta=size, y=eggs16] {\dataNumericC};
        \addplot[style={rred,fill=rred,mark=none}] table [meta=size, y=us16] {\dataNumericC};
        \addplot[style={ggreen,fill=ggreen,mark=none}] table [meta=size, y=gpu] {\dataNumericC};
 
        \legend{EGGS (1 thread), {ours \hspace{7.25pt} (1 thread)}, EGGS (16 threads), {ours \hspace{7.25pt} (16 threads)}, ours \hspace{8.25pt} (\device{Vega 20})}
    \end{axis}
\end{tikzpicture}
\end{center}
		\vspace{12pt}
	\end{center}
	\def\arraystretch{1.15}
	\begin{center}
	\begin{tabular}{rccc}
		                                                                          & 100k    & 500K     & 1M       \\\toprule
		$(\alpha \mathbf A + \mathbf B)^\top (\beta \mathbf B^\top + \mathbf C) $ & 1s 225ms & 8s 730ms  & 19s 514ms \\
		$\mathbf A \mathbf B \mathbf C $                                          & 1s 803ms & 11s 400ms & 25s 408ms \\
		$(\mathbf A + \mathbf B) (\mathbf A \mathbf B + \mathbf C) $              & 1s 528ms & 10s 549ms & 23s 938ms \\\hline
	\end{tabular}
	\end{center}
	\def\arraystretch{1.}
	\caption{We compare the speedups of our method and EGGS for the evaluation of sparse algebraic expressions. The reference timings (on \device{AMD}) of a direct Eigen implementation are provided in the table.}
	\label{fig:sparseSpeedup}
\end{figure}
We compare our method with EGGS~\cite{eggs} by repeating an experiment in that paper. We generate code to evaluate three arithmetic expressions using 32-bit floats involving sparse matrices $\mathbf A, \mathbf B, \mathbf C \in \R^{n\times n}$ for $n =$ 100k, 500k, and 1M. All sparse matrices are generated by randomly choosing 6 non-zero elements per row. Both methods generate code based on operations of sparse Eigen matrices with custom template parameters for the numeric type.
For this basic experiment, our method is very similar to EGGS because the expressions are relatively easy, and storing intermediate results is not necessary.
All methods outperform Eigen by orders of magnitude. Our approach is faster than EGGS in all examples (Figure~\ref{fig:sparseSpeedup}), which can be attributed to the optimized code generation and memory organization.

\subsection{Structured sparse products} \label{sec:structured}
Randomized sparse matrices are a special case.
When computing a simple sparse matrix product $\mathbf R = \mathbf A \mathbf B$, almost all non-zero values of the resulting matrix are computed by multiplying just a pair of matrix elements, one from $\mathbf A$ and another from $\mathbf B$.
If matrices are structured, for example based on a triangle mesh like the cotan Laplacian, the situation is different.
Each value in $\mathbf R$ will be the sparse inner product of two vectors that share more than one non-zero position.
Consequently, more kernels will be generated by our approach. If we now consider the triple product $\mathbf R = \mathbf A \mathbf B \mathbf C$, the number of kernels will be even larger due to the different combinations these inner products can interact with elements of $\mathbf C$. Considering even more factors leads to combinatoric explosion. Our method circumvents this issue by identifying common subexpressions, which will include most of the results computed while performing the product of $\mathbf A$ and $\mathbf B$ in this example.
To demonstrate this feature, we compute the matrix power $\mathbf L^3$ of the cotan operator $\mathbf L$ for meshes with $n = 20k, 50k$ and $200k$ vertices. In all cases, our method generates only 22 kernels while EGGS generates more than $10^3$, with even larger numbers for $n=50k$ and $200k$. This leads to excessive compilation times and slow execution performance.

\hide{
The following table breaks down the time spent by our method in different stages of the pipeline on \device{AMD}. 
\vspace{2pt}
\begin{center}
	\begin{tabular}{lrrrrr}
		n    & execution & analyze    & generate & compile \\\toprule
		10k  & 1s 862ms   & 13s 127ms  & 239ms    & 1s 573ms \\
		50k  & 8s 990ms   & 1m 4s 190ms  & 1s 057ms  & 1s 687ms \\
		200k & 36s 168ms  & 4m 28s 414ms & 4s 856ms  & 1s 542ms
	\end{tabular}%
\vspace{2pt}\\%
\end{center}%
The execution phase runs the original program with symbolic types.
The analysis includes expression decomposition and grouping, leaf harvesting, and expression optimization. Code generation writes the code file and constructs all position indices for indirect memory access. Compilation creates the final program based on the generated code (we use gcc in this case). Most time is spent analyzing the expressions and their dependencies since all subexpressions above the complexity threshold $\cc$ have to be considered, indexed, and eventually decomposed.}

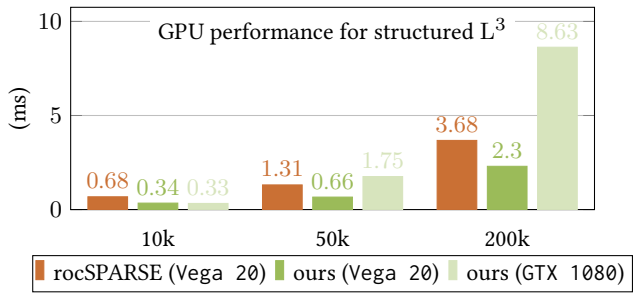
\begin{figure}[t!]
	\pgfplotstableread{
size  hip vega gtx
10k 0.67967 0.34401 0.3261 
50k 1.3109 0.66347 1.75414 
200k 3.67561 2.29999 8.6286
}\LLLData

\begin{center}
\begin{tikzpicture}
    \begin{axis}[
        title={GPU performance for structured $\mathbf L^3$},
        title style={at={(0.5,0.91)},
                   anchor=north, fill=white},
        width  = \linewidth,
        height = 0.5*\linewidth,
        major x tick style = transparent,
        ybar=4pt,
        bar width=15pt,
        ymajorgrids = true,
        symbolic x coords={10k, 50k, 200k},
        xtick = data,
        scaled y ticks = false,
        enlarge x limits=0.25,
        enlarge y limits=0.,
        ymin=0,
        ymax=10.75,
        nodes near coords,
        ylabel={\normalsize (ms)},
        ylabel style={yshift=-0.55cm},
          legend columns = 3,
        legend style = {at={(0.5, -0.225)}, anchor=north, inner sep=1pt, style={column sep=0.05cm}},       
        legend cell align=left,
    ]
        \addplot[ style={bbrown,fill=bbrown,mark=black}
        ] table [meta=size,y=hip] {\LLLData};

          \addplot[style={ggreen,fill=ggreen,mark=black}
        ] table [meta=size,y=vega] {\LLLData};

         \addplot[style={ggreen!40,fill=ggreen!40,mark=black}
        ] table [meta=size,y=gtx] {\LLLData};
       \legend{rocSPARSE (\device{Vega 20}), ours (\device{Vega 20}), ours (\device{GTX 1080})}
    \end{axis}
\end{tikzpicture}
\end{center}
	\caption{We compare the performance of our method on different GPUs with the performance of a dedicated linear algebra library targeted at GPUs.}
	\label{fig:LLL}
\end{figure}
In \figref{fig:LLL} we compare the performance of computing the matrix power $\mathbf L^3$ of our method on different GPUs with the performance of a dedicated linear algebra library targeted at GPUs \cite{rocmsparse}. The library function analyses both products $\mathbf L^2 = \mathbf L \mathbf L$ and $\mathbf L^3 = \mathbf L^2 \mathbf L$ in a preprocess and tries to find an optimal computation strategy (not included in the timings). Our method is faster on the same GPU while being far more general. All methods are orders of magnitude faster than Eigen and CPU versions of our code. The performance of our code is better on \device{Vega 20} due to its higher number of ALUs.

\subsection{Cotan operator construction}
\label{sec:cotanExample}
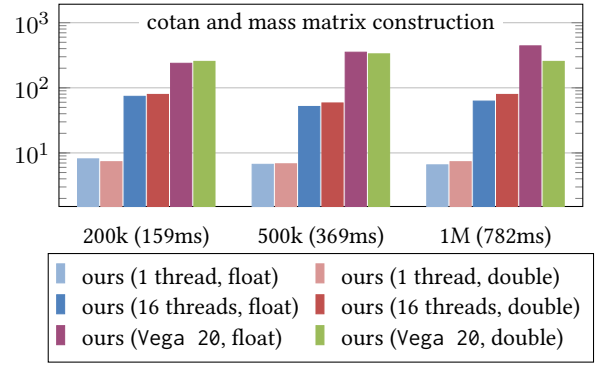
\begin{figure}[t!]
	\pgfplotstableread{
size  us1  us16  gpu             us1d  us16d  gpud   
200k 8.1 74 238                  7.32659 79.1186 254.547   
500k 6.68027 51.6796 352.311     6.80367 58.5247 332.405 
1M   6.55781 62.534 440.9        7.32659 79.1186 254.547
}\cotanData
\begin{tikzpicture}
    \begin{axis}[
        title={{cotan and mass matrix construction}},
        title style={fill=white, at={(0.5,0.92)},  anchor=north},
        width  = 1.\linewidth,
        height = 0.5*\linewidth,
        major x tick style = transparent,
        ybar=2*\pgflinewidth,
        ymode=log,
        bar width=8pt,
        ymajorgrids = true,
        symbolic x coords={200k,500k,1M},
        scaled y ticks = false,
        enlarge x limits=0.25,
        enlarge y limits=0.35,
        ymin=0,
        legend columns = 2,
        legend style = {at={(0.5, -0.23)}, anchor=north, inner sep=3pt, style={column sep=0.15cm}},       
        legend cell align=left,
        xtick = data,
        xticklabels={{200k (159ms)}, {500k (369ms)}, {1M (782ms)}}
    ]
        \addplot[style={bblue!60,fill=bblue!60}] table [meta=size, y=us1] {\cotanData};
        \addplot[style={rred!60,fill=rred!60,mark=none}] table [meta=size, y=us1d] {\cotanData};

        \addplot[style={bblue,fill=bblue}] table [meta=size, y=us16] {\cotanData};
        \addplot[style={rred,fill=rred,mark=none}] table [meta=size, y=us16d] {\cotanData};

        \addplot[style={ppurple,fill=ppurple,mark=none}] table [meta=size, y=gpu] {\cotanData};
        \addplot[style={ggreen,fill=ggreen,mark=none}] table [meta=size, y=gpud] {\cotanData};
    
        \legend{{ours (1 thread, float)},{ours (1 thread, double)},
                {ours (16 threads, float)},{ours (16 threads, double)}, 
                {ours (\device{Vega 20}, float)}, {ours (\device{Vega 20}, double)}}
    \end{axis}
\end{tikzpicture}
	\caption{Speedups with respect to a standard implementation using 32- as well as 64-bit floating-point accuracy. For meshes with many vertices, the construction of the cotan Laplacian is computationally expensive. We report the timings of the original code in parenthesis. CPU timings in this figure were measured on the \device{AMD} processor.}
	\label{fig:cotan}
\end{figure}%
Constructing the cotan Laplacian is a frequent operation in geometry processing. For three meshes of 200k, 500k, and 1M vertices, we generate code for the assembly of the sparse Laplacian and the corresponding barycentric mass matrix. We report speedups for float and double versions of the code in Figure~\ref{fig:cotan}. 
Constructing the operator proceeds in two phases. First, contributions for each face are computed and added to a list of \texttt{(row, column, value)} triplets. This triplet list is then used to construct the final sparse matrix.
Since the final step takes up a significant amount of time, we report reference timings for the construction of the triplet list alone.
\vspace{5pt}
\begin{center}
	\begin{tabular}{rccc}\toprule
		mesh size    & 200k  & 500k  & 1M    \\\hline
			compute triplets only & 64$\,$ms  & 112$\,$ms & 223$\,$ms\\
    		total time & 159$\,$ms & 396$\,$ms & 782$\,$ms \\\bottomrule
	\end{tabular}%
\end{center}%
\vspace{5pt}%
Because the structure of the sparse matrix is known at compile time for our method, we can generate code that does not suffer from this performance bottleneck.
\subsubsection{Comparison to EGGS}
For a mesh with 200k vertices we compare our method to EGGS. Since EGGS can not decompose the expressions for each matrix element, the generated code will compute a lot of redundant subtrees. 
\vspace{5pt}
\begin{center}
\begin{tabular}{lccccc}\toprule
& reference  & \multicolumn{2}{c}{EGGS} & \multicolumn{2}{c}{ours} \\\hline
\#threads&1   & 1        & 16        & 1         & 16         \\\hline
time &159ms & 165ms       & 25ms       & 22ms       & 1.5ms\\\bottomrule   
\end{tabular}
\end{center}
\vspace{5pt}%
As a consquence, EGGS' single-threaded performance is worse than the reference implementation. Our method automatically identifies redundant subtrees and computes them only once resulting in a 6$\times$ speedup (100 $\times$ when using 16 threads). The difference can be even more pronounced for more complex examples.

\begin{figure}[t!]
	\pgfplotstableread{
size ref1 ref2 us1      us16    gpu 
7k   25   57   11.9     2.5     3.855
20k  92   205  37.8     6.4     4.884
100k 686  1277 307.4    39.3    7.862
}\dataDualLaplace

\begin{tikzpicture}
    \begin{axis}[
        title={{volumetric dual Laplace and mass matrix}},
        title style={fill=white, at={(0.5,0.92)},  anchor=north},
        width  = 1.\linewidth,
        height = 0.5*\linewidth,
        major x tick style = transparent,
        ybar=2*\pgflinewidth,
        bar width=8pt,
        ymode=log,
        ymajorgrids = true,
        symbolic x coords={7k,20k,100k},
        scaled y ticks = false,
        enlarge x limits=0.25,
        ymin=0,
        ymax=5000,
        legend columns = 2,
        legend style = {at={(0.5, -0.23)}, anchor=north, inner sep=3pt, style={column sep=0.15cm}},       
        legend cell align=left,
        xtick = data,
        xticklabels={{7k}, {20k}, {100k}},
        ylabel={\normalsize (ms)},
        ylabel style={yshift=-0.48cm},
    ]
        \addplot[style={ppurple,fill=ppurple}] table [meta=size, y=ref2] {\dataDualLaplace};
        \addplot[style={ppurple!60,fill=ppurple!60,mark=none}] table [meta=size, y=ref1] {\dataDualLaplace};

        \addplot[style={rred!60,fill=rred!60}] table [meta=size, y=us1] {\dataDualLaplace};
        \addplot[style={rred,fill=rred,mark=none}] table [meta=size, y=us16] {\dataDualLaplace};

        \addplot[style={ggreen,fill=ggreen,mark=none}] table [meta=size, y=gpu] {\dataDualLaplace};
    
        \legend{{original},{original (values only)},
                {ours (1 thread)},{ours (16 threads)}, 
                {ours (\device{Vega 20})}}
    \end{axis}
\end{tikzpicture}
	\caption{Construction of the dual volumetric Laplacian and mass matrix \cite{Alexa:PLOTM:2020}. We compare to the implementation provided by the authors both, with and without final matrix assembly. All timings in this figure were taken on \device{AMD}.}
	\label{fig:dualLaplace}
\end{figure}
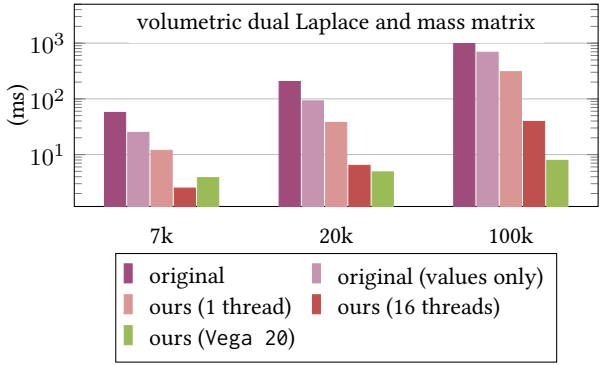
\subsection{Volumetric Laplacian}
Recently Alexa et al.~\shortcite{Alexa:PLOTM:2020} investigated Laplace operators for tetrahedral meshes and provided code for the construction of the volumetric dual Laplacian.
We replaced all numeric types in the provided code and generate code for the construction of the Laplacian and Voronoi mass matrix for several tetrahedral meshes (7k, 20k, and 200k vertices).
The code contains the computation of $3\times 3$ determinants and dense matrix inversions which leads to more complex expressions compared to the cotan operator example in the previous section.
We report absolute timings in milliseconds and compare against the construction of the triplets with and without full sparse matrix construction (Figure~\ref{fig:dualLaplace}). In all cases, our method is order(s) of magnitude faster. We observe that the GPU implementation provides performance benefits only if enough kernel instances (meshes with >7k vertices) are executed.

\subsection{As-rigid-as-possible surface modeling}
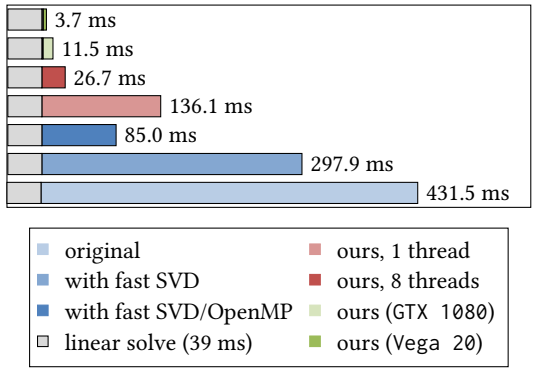
\begin{figure}[t!]
	\begin{tikzpicture}
    \begin{axis}[
    xbar stacked,   
    xmin=0,         
    xmax=600,
    ymin=0.5,
    ymax=7.5,
    bar width = 8pt,
    width=1\linewidth,
    height=0.5\linewidth,
    ticks=none,
    legend columns = 2,
    legend style = {at={(0.5, -0.1)}, anchor=north, inner sep=3pt, style={column sep=0.15cm}},       
    legend cell align=left,  
      legend entries={ original, {ours, 1 thread}, {with fast SVD}, {ours, 8 threads},  {with fast SVD/OpenMP}, {ours (\device{GTX 1080})}, {linear solve (39 ms)}, 
      {ours (\device{Vega 20})}}
   ]
    \addlegendimage{only marks, mark=square*, mark options={draw=bblue!40,fill=bblue!40}}
    \addlegendimage{only marks, mark=square*, mark options={solid,draw=rred!60,fill=rred!60}}
    \addlegendimage{only marks, mark=square*, mark options={solid,draw=bblue!70,fill=bblue!70}}
    \addlegendimage{only marks, mark=square*, mark options={solid,draw=rred,fill=rred}}
    \addlegendimage{only marks, mark=square*, mark options={solid,draw=bblue,fill=bblue}}
    \addlegendimage{only marks, mark=square*, mark options={draw=ggreen!40,fill=ggreen!40}}
    \addlegendimage{only marks, mark=square*, mark options={draw=black,fill=black!15}}
    \addlegendimage{only marks, mark=square*, mark options={solid,draw=ggreen,fill=ggreen}}

    \addplot[fill=black!15] coordinates {(39, 1)};
    \addplot[fill=bblue!40] coordinates {(431.490, 1)} node[right,pos=0.5]{431.5 ms};
    
    \resetstackedplots
    \addplot[fill=black!15] coordinates {(39, 2)};
    \addplot[fill=bblue!70] coordinates {(297.9, 2)} node[right,pos=0.5]{297.9 ms};

    \resetstackedplots
    \addplot[fill=black!15] coordinates {(39, 3)};
    \addplot[fill=bblue] coordinates {(85.0, 3)} node[right,pos=0.5]{85.0 ms};

    \resetstackedplots
    \addplot[fill=black!15] coordinates {(39, 4)};
    \addplot[fill=rred!60] coordinates {(136.1, 4)} node[right,pos=0.5]{136.1 ms};

    \resetstackedplots
    \addplot[fill=black!15] coordinates {(39, 5)};
    \addplot[fill=rred] coordinates {(26.7, 5)} node[right,pos=0.5]{26.7 ms};

    \resetstackedplots
    \addplot[fill=black!15] coordinates {(39, 6)};
    \addplot[fill=black!25] coordinates {(1.2, 6)};   
    \addplot[fill=ggreen!40] coordinates {(11.500, 6)} node[right,pos=0.5]{11.5  ms};
   
    \resetstackedplots
    \addplot[fill=black!15] coordinates {(39, 7)};
    \addplot[fill=black!25] coordinates {(1.48, 7)};   
    \addplot[fill=ggreen] coordinates {(3.73, 7)} node[right,pos=0.5]{3.7  ms};

    \end{axis}
\end{tikzpicture}
	\caption{An iteration of as-rigid-as-possible surface deformation consists of the construction of a right hand side that is subsequently used to solve a linear system. Optimizing code by hand can significantly improve performance. We automatically generate optimized code that is even more efficient. All experiments in this figure were conducted on \device{Intel}.}
	\label{fig:arapRHS}
\end{figure}
As-rigid-as-possible surface modeling is an iterative method for deforming triangle meshes \cite{arap}. The algorithm proceeds by alternating two steps. In the local step, rotations are estimated for each vertex which involves the singular value decomposition (SVD) of $3\times 3$ matrices.  The local rotations are used to build a dense matrix $\mathbf B \in \R^{n\times 3}$ that is subsequently used to solve the linear system $\mathbf L \mathbf X = \mathbf B$ where $\mathbf L$ represents the cotan operator of the initial mesh. Since the system matrix is fixed as long as the boundary constraints of the deformation problem do not change, we can precompute a factorization. The deformation loop boils down to the construction of the right-hand side $\mathbf B$ and a linear solve by means of forward- and back substitution. We generate optimized code for the local step, including the construction of $\mathbf B$. Directly translating the code poses a problem since the SVD based on Jacobi rotations, as implemented in Eigen, contains a loop that checks for a convergence criterion. We adopt the strategy of the heavily optimized SVD implementation by McAdams et al.~\shortcite{McAdams:2011} (fast SVD) and fix the loop count to $4$. Furthermore, we mark the input and output of the SVD computation as explicit intermediate values (detailed in \secref{sec:explicitIntermediates}), so that the full SVD computation is executed in a single kernel.

In \figref{fig:arapRHS} we compare to the original implementation as well as to an optimized version that employs the hand-vectorized optimized implementation of McAdams et al.~\shortcite{McAdams:2011} and a version that has been additionally annotated with OpenMP's \texttt{\#pragma omp parallel for} to parallelize inner loops. These hand optimizations can significantly improve performance but are still outperformed by our code. For the experiment we used the armadillo mesh with 300k vertices. Due to the local nature of the algorithm, the timings scale almost linearly with mesh size.
For reference, we report the time taken to perform forward- and back substitution for the three columns of $\mathbf b$ using Pardiso \cite{pardiso} (8 threads, Intel MKL).
By using our method, it is possible to shift the bottleneck from the local step to the the linear system solve. The GPU implementation is able to reduce the computational load of this step to less than $10\%$ of the linear solve. The GPU timings include the memory transfer of the current geometry to the device and the transfer of the dense matrix $\mathbf b$ back to the host, which accounts for roughly one-third of the \unit[3.7]{ms}. The GPU is particularly useful here since the amount of computation is relatively high compared to the amount of data transferred. Our method enables real-time interactive shape deformation for this relatively large mesh by increasing performance from \unit[2.1]{fps} to \unit[23.4]{fps}.

\begin{table*}[!t]
\begin{tabular}{lllllllllll}
\toprule
 & \multicolumn{1}{c}{\begin{tabular}[c]{@{}c@{}}Eigen\\ \device{Intel}\end{tabular}}  &
   \multicolumn{1}{c}{\begin{tabular}[c]{@{}c@{}}Taco\\ \device{Intel}\end{tabular}} &
   \multicolumn{1}{c}{\begin{tabular}[c]{@{}c@{}}MKL\\ \device{Intel}\end{tabular}} & \multicolumn{1}{c}{\begin{tabular}[c]{@{}c@{}}MKL$\dagger$\\ \device{Intel}\end{tabular}} & \multicolumn{1}{c}{\begin{tabular}[c]{@{}c@{}}ours$\dagger$\\ \device{Intel}\end{tabular}} & \multicolumn{1}{c}{\begin{tabular}[c]{@{}c@{}}rocSparse$\dagger$\\ \device{Vega}\end{tabular}} & \multicolumn{1}{c}{\begin{tabular}[c]{@{}c@{}}ours\\ \device{Vega}\end{tabular}} & \multicolumn{1}{c}{\begin{tabular}[c]{@{}c@{}}cuSPARSE$\dagger$\\ \device{GTX}\end{tabular}} &  \multicolumn{1}{c}{\begin{tabular}[c]{@{}c@{}}ours$\dagger$\\ \device{GTX}\end{tabular}} \\ \hline
$\mathbf A^2$ & 36ms & \revv{1.9ms}& 3.1ms & 1.57ms & 0.62ms & 0.35ms & 0.17ms & 0.23ms &  0.28ms \\
$\mathbf A^3$& 106ms & \revv{5.8ms}& 10.6ms & 5.2ms & 3.98ms & 1.31ms & 0.49ms & 0.52ms & 1.03ms \\
$\mathbf A^4$ & 193ms & \revv{13.4ms}& 21.8ms & 14.7ms & 13.0ms & 2.23ms & 0.98ms & 1.54ms &  2.39ms\\
$\mathbf L \mathbf M \mathbf L^\top + \mathbf K$ &  1s36ms & 261ms & 464ms & 197ms & 156ms & 13.6ms & 3.89ms  & - & 16.8ms\\
\bottomrule
\end{tabular}
\caption{We compute powers of a sparse matrix $\mathbf A \in \mathds R^{50k \times 50k}$ and evaluate the sparse expression $\mathbf L \mathbf M \mathbf L^\top + \mathbf K$ with $\mathbf L, \mathbf M, \mathbf K \in \mathds R^{500k \times 500k}$. The matrices $\mathbf A$ and $\mathbf L$ are mesh Laplacians, $\mathbf M$ is a mass matrix and $\mathbf K$ contains six random non-zeros per row. We compute timings using
  several linear algebra packages on different devices. The $\dagger$ symbol indicates that the method uses problem specific precomputation like preallocation and scheduling (not included in the timings). Our method is competitive even though it is not specialized to matrix-matrix products in contrast to all other presented options.
}
\label{tab:Lpow}
\end{table*}
\subsection{Cloth simulation}\label{sec:cloth}
\begin{figure}[t!]
	\begin{tikzpicture}[  
    every axis/.style={ybar stacked,
    ymin=0,ymax=5.7,
    ymajorgrids = true,
    symbolic x coords={20k,50k,100k},
    scaled y ticks = false,
    enlarge x limits=0.25,
    bar width=8pt,
     major x tick style = transparent,
    }
]

\begin{axis}[bar shift=-15pt, bar width=8pt,
    title={{cloth simulation: system construction}},
    title style={fill=white, at={(0.55,0.96)},  anchor=north},
    ytick={0,1,2,4,6},
    xtick = data,
    xticklabels={{20k}, {50k}, {100k}},
    ylabel={\normalsize performance relative to factorization},
    ylabel style={yshift=-0.7cm},
    legend columns = 3,
    legend style = {at={(0.5, -0.1)}, anchor=north, inner sep=3pt, style={column sep=0.15cm}},       
    legend cell align=left,  
    legend entries={{factorization}, {linear solve}, {reference}, {ours (1 thread)}, {ours (8 threads)}, {ours (\device{GTX 1080})}, {ours (\device{Vega 20})}}
]

    \addlegendimage{only marks, mark=square*, mark options={draw=black!60,fill=black!60, yshift=-5pt}}
    \addlegendimage{only marks, mark=square*, mark options={solid,draw=black!40,fill=black!40, yshift=-5pt}}
    \addlegendimage{only marks, mark=square*, mark options={solid,draw=ppurple,fill=ppurple, yshift=-5pt}}
    \addlegendimage{only marks, mark=square*, mark options={solid,draw=rred!60,fill=rred!60, yshift=-5pt}}
    \addlegendimage{only marks, mark=square*, mark options={solid,draw=rred,fill=rred, yshift=-5pt}}
    \addlegendimage{only marks, mark=square*, mark options={solid,draw=ggreen!40,fill=ggreen!40, yshift=-5pt}}
    \addlegendimage{only marks, mark=square*, mark options={solid,draw=ggreen,fill=ggreen, yshift=-5pt}}

\addplot[fill=black!60, draw=black!60] coordinates
{(20k,1) (50k,1) (100k,1)};
\addplot[fill=black!40, draw=black!40] coordinates
{(20k, 0.20138) (50k, 0.153125) (100k, 0.112168)};
\addplot[fill=ppurple, draw=ppurple] coordinates
{(20k, 5.7261) (50k, 4.13307)  (100k, 3.47251)};
\end{axis}

\begin{axis}[hide axis, bar shift=-25pt]
\addplot+[fill=none, draw=black!60] coordinates
{(20k,1) (50k,1) (100k,1)};
\end{axis}

\begin{axis}[hide axis, bar shift=-5]
\addplot+[fill=black!60, draw=black!60] coordinates
{(20k,1) (50k,1) (100k,1)};
\addplot+[fill=black!40, draw=black!40] coordinates
{(20k, 0.20138) (50k, 0.153125) (100k, 0.112168)};
\addplot+[fill=rred!60,draw=rred!60] coordinates
{(20k, 0.709) (50k, 0.666227) (100k, 0.709482)};
\end{axis}

\begin{axis}[hide axis,bar shift=5pt]
\addplot+[fill=black!60, draw=black!60] coordinates
{(20k,1) (50k,1) (100k,1)};
\addplot+[fill=black!40, draw=black!40] coordinates
{(20k, 0.20138) (50k, 0.153125) (100k, 0.112168)};
\addplot+[fill=rred,draw=rred] coordinates
{(20k, 0.296597) (50k, 0.325626) (100k, 0.273058)};
\end{axis}

\begin{axis}[bar shift=15pt, hide axis]
\addplot+[fill=black!60, draw=black!60] coordinates
{(20k,1) (50k,1) (100k,1)};
\addplot+[fill=black!40, draw=black!40] coordinates
{(20k, 0.20138) (50k, 0.153125) (100k, 0.112168)};
\addplot+[fill=ggreen!40,draw=ggreen!40] coordinates
{(20k, 0.077) (50k, 0.064) (100k, 0.051)};
\end{axis}

\begin{axis}[bar shift=25pt, hide axis]
\addplot+[fill=black!60, draw=black!60] coordinates
{(20k,1) (50k,1) (100k,1)};
\addplot+[fill=black!40, draw=black!40] coordinates
{(20k, 0.20138) (50k, 0.153125) (100k, 0.112168)};
\addplot+[fill=ggreen,draw=ggreen] coordinates
{(20k, 0.07) (50k, 0.0344223) (100k, 0.0197197)};
\end{axis}

\node at (1.25,4.8) {$\leftarrow 5.7$};
\node [rotate=90, anchor=north west] at (0.06, 0.02) {\footnotesize{49 ms}};
\node [rotate=90, anchor=north west] at (2.35, 0.02) {\footnotesize{156 ms}};
\node [rotate=90, anchor=north west] at (4.65, 0.02) {\footnotesize{553 ms}};



\end{tikzpicture}
	\caption{Constructing the linear systems can dominate runtime for cloth simulation. We report the performance of the original code and versions generated using our system relative to the time it takes to factorize the matrix. Our method removes the bottleneck. Timings taken on \device{Intel}.}
	\label{fig:clothTimes}
\end{figure}
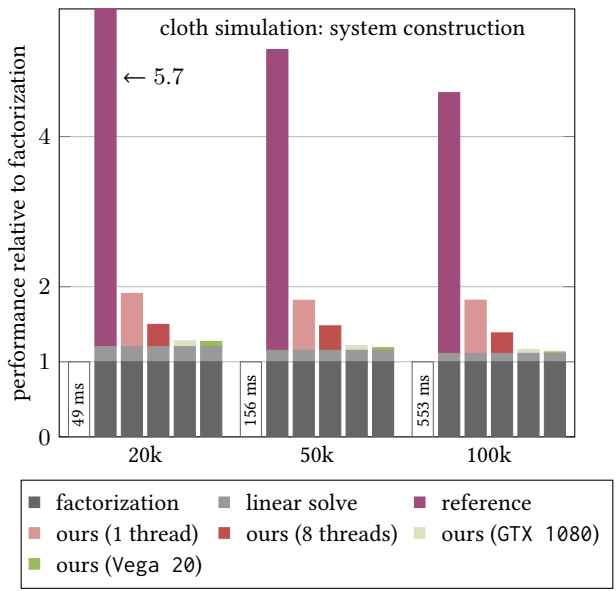
\begin{figure}[t!]
	\includegraphics[width=\linewidth]{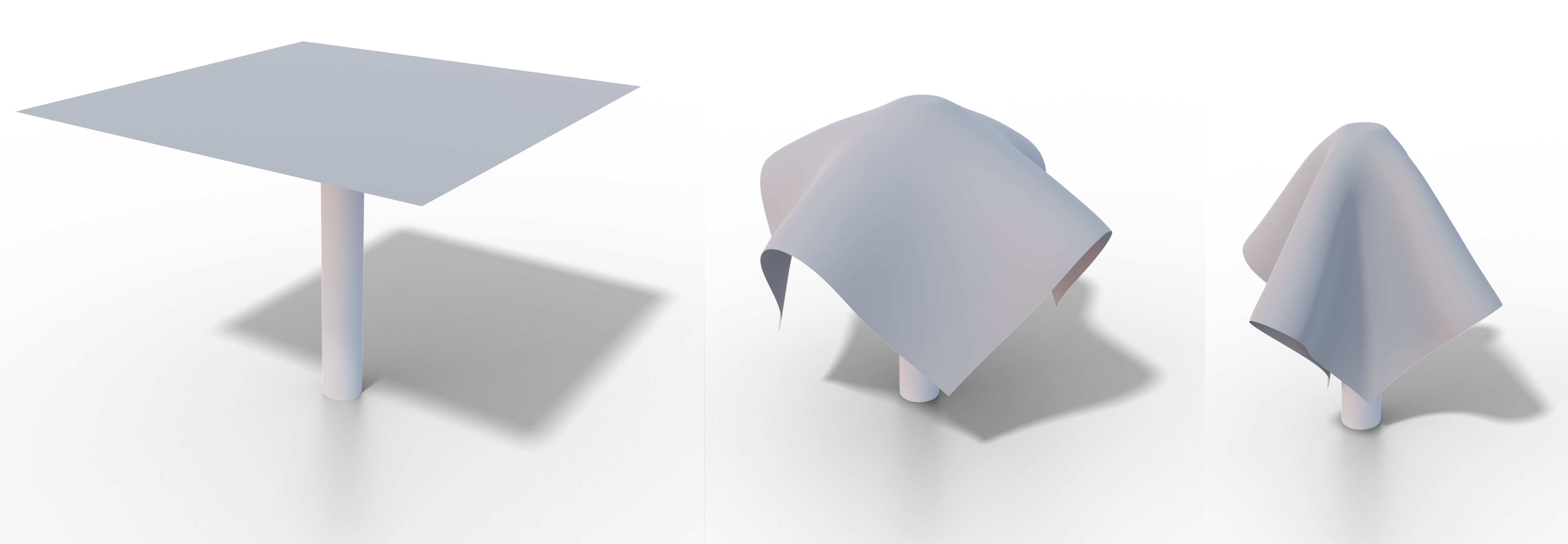}
	\caption{Several frames of a cloth simulation using the method of Baraff and Witkin~\shortcite{Baraff:1994} as implemented by Wolff et al.~\shortcite{Wolff:2021}.}
	\label{fig:cloth}
\end{figure}
The classic cloth simulation method by Baraff and Witkin~\shortcite{Baraff:1994} has been extended in several ways over the years and still enjoys popularity today.
The method defines stretching, shearing, and bending energies that lead to damping and direct forces. The method is implicit and solves a linear system at each time step.
However, the majority of the time is spent constructing the linear system, which poses a performance bottleneck.
To demonstrate the practicality of our system for more complex computations, we generate code for assembling the system matrix and the right-hand side. To this end we run a templated version of an implementation provided by Wolff et al.\ \shortcite{Wolff:2021}.
The final system has the form
\begin{align*}
	\mathbf A = \mathbf M & - h  (\mathbf D_{\mathrm{stretch}} + \mathbf D_{\mathrm{shear}} + \mathbf D_{\mathrm{bend}}) \\ &+ h^2 (\mathbf K_{\mathrm{stretch}} + \mathbf K_{\mathrm{shear}} + \mathbf K_{\mathrm{bend}})\\
\end{align*}
where $\mathbf M$ is the mass matrix and $\mathbf K$ and $\mathbf D$ are stiffness and damping matrices, respectively. 
All matrices are constructed by combining dense per-element contributions and we tag these blocks for the stretch and shear energies explicitly (see \secref{sec:explicitIntermediates}). 
Computing the local Hessians of the stretch and shear energy takes 587ms for a mesh with 100k vertices using our reference implementation. Generating code without tagging improves runtime by a factor of 5 (104ms) while enabling the feature gives a factor of 40 (14.7ms) in speedup. All three measurements have been conducted on \device{Intel} using a single thread.
In \figref{fig:clothTimes} we compare the performance of our method with the original implementation.
All timings are measured on our \device{Intel} machine and are reported relative to the time it takes to factor the system matrix (Pardiso, 8 threads, Intel MKL). With increasing mesh size, the relative time spent on factorization increases. In all cases, we can significantly improve the performance and remove the bottleneck of system construction. The GPU timings include data transfer from and to the device; thus, the system construction becomes a negligible part of the overall runtime.

\subsubsection{Automatic differentiation of deformation energies}\label{sec:autodiff-comp}
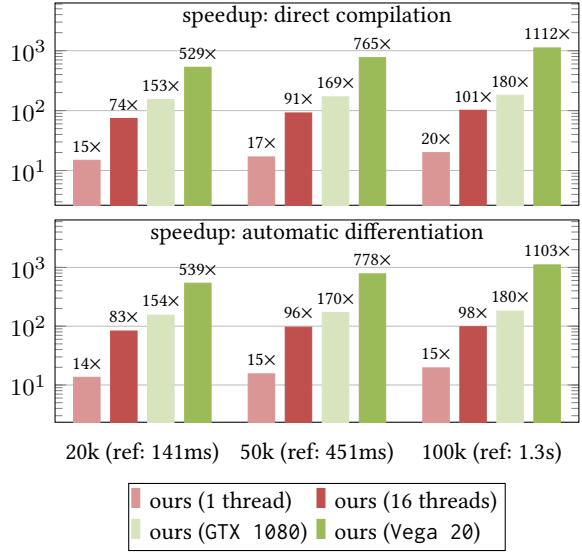
\begin{figure}[t!]
	\pgfplotstableread{
size    us1     us16    gpu   gpuCuda labelCuda  labelus1        labelus16       labelgpu
20k 14.8    74. 528.571      152.628    {153$\times$}     {15$\times$}    {74$\times$}    {529$\times$}
50k 16.8321 91.3158 764.576   169.496    {169$\times$}         {17$\times$}    {91$\times$}    {765$\times$}
100k    19.877  101.386 1112.  180.214    {180$\times$}       {20$\times$}    {101$\times$}   {1112$\times$}
}\clothHessianDirect
\pgfplotstableread{
size  us1  us16  gpu         gpuCuda  labelCuda         labelus1        labelus16       labelgpu
20k 13.5192 82.7059 538.697  154.26    {154$\times$}         {14$\times$}    {83$\times$}    {539$\times$}
50k 15.4486 96.3889 777.759  169.873    {170$\times$}               {15$\times$}    {96$\times$}    {778$\times$}   
100k 19.5472 97.8873 1103.17 180.274    {180$\times$}         {15$\times$}    {98$\times$}    {1103$\times$}
}\clothHessianAD
\begin{center}
\begin{tikzpicture}
    \begin{axis}[
        title={speedup: direct compilation},
        title style={at={(0.5,0.95)},
                   anchor=north},
        width  = \linewidth,
        height = 0.5*\linewidth,
        major x tick style = transparent,
        ybar=4pt,
        ymode=log,
        bar width=10pt,
        ymajorgrids = true,
        symbolic x coords={20k, 50k, 100k},
        xtick = \empty,
        scaled y ticks = false,
        enlarge x limits=0.25,
        enlarge y limits=0.4,
        ymin=0,
        nodes near coords,
    ]
        \addplot[ nodes={font=\footnotesize,black},
            point meta=explicit symbolic,
            style={rred!60,fill=rred!60,mark=black}
        ] table [meta=labelus1,y=us1] {\clothHessianDirect};

         \addplot[ nodes={font=\footnotesize,black},
            point meta=explicit symbolic,
            style={rred,fill=rred,mark=black}
        ] table [meta=labelus16,y=us16] {\clothHessianDirect};

          \addplot[ nodes={font=\footnotesize,black},
            point meta=explicit symbolic,
            style={ggreen!40,fill=ggreen!40,mark=black}
        ] table [meta=labelCuda,y=gpuCuda] {\clothHessianDirect};

         \addplot[ nodes={font=\footnotesize,black},
            point meta=explicit symbolic,
            style={ggreen,fill=ggreen,mark=black}
        ] table [meta=labelgpu,y=gpu] {\clothHessianDirect};

    \end{axis}
\end{tikzpicture}
\begin{tikzpicture}
    \begin{axis}[
        title={speedup: automatic differentiation},
        title style={at={(0.5,0.95)},
                   anchor=north},
        width  = \linewidth,
        height = 0.5*\linewidth,
        major x tick style = transparent,
        ybar=4,
        ymode=log,
        bar width=10pt,
        ymajorgrids = true,
        symbolic x coords={20k, 50k, 100k},
        xtick = data,
        xticklabels={{20k (ref: 141ms)}, {50k (ref: 451ms)}, {100k (ref: 1.3s)}},
        scaled y ticks = false,
        enlarge x limits=0.25,
        enlarge y limits=0.4,
        ymin=0,
        nodes near coords,
        legend columns = 2,
        legend style = {at={(0.5, -0.3)}, anchor=north, inner sep=1pt, style={column sep=0.05cm}},       
        legend cell align=left,
    ]
        \addplot[ nodes={font=\footnotesize,black},
            point meta=explicit symbolic,
            style={rred!60,fill=rred!60,mark=black}
        ] table [meta=labelus1,y=us1] {\clothHessianAD};

         \addplot[ nodes={font=\footnotesize,black},
            point meta=explicit symbolic,
            style={rred,fill=rred,mark=black}
        ] table [meta=labelus16,y=us16] {\clothHessianAD};

            \addplot[ nodes={font=\footnotesize,black},
            point meta=explicit symbolic,
            style={ggreen!40,fill=ggreen!40,mark=black}
        ] table [meta=labelCuda,y=gpuCuda] {\clothHessianAD};

         \addplot[ nodes={font=\footnotesize,black},
            point meta=explicit symbolic,
            style={ggreen,fill=ggreen,mark=black}
        ] table [meta=labelgpu,y=gpu] {\clothHessianAD};

       \legend{ours (1 thread), ours (16 threads), ours (\device{GTX 1080}), ours (\device{Vega 20})}
    \end{axis}
\end{tikzpicture}
\end{center}
	\caption{Our system offers two options to efficiently compute the Hessian and gradient of a deformation energy \cite{Baraff:1994}: We can either directly compile optimized code based on an existing implementation, or employ automatic differentiation based on a symbolic representation of the energy integrated over the mesh. Both versions are competitive and provide significant speedups over the original implementations (reference times in parenthesis). Timings taken on \device{AMD}.}
	\label{fig:clothAD}
\end{figure}
Several parts of the system matrix and its right-hand side are constructed from gradients and Hessians of deformation energies.
To demonstrate our symbolic auto differentiation feature, we focus on the sum of stretch and shear energies and compute their gradient and Hessian.
This gives us two methods to generate code computing derivatives for a specific mesh: (1) We can directly generate derivative expressions based on the implemented energy using automatic differentiation or (2) we can use run the hand optimized code for constructing gradient and Hessian using our symbolic type just as in the previous experiment. In all cases, we achieve at least an order of magnitude performance increase. Interestingly, both versions are competitive with a slight benefit for direct compilation of hand-optimized code when using a single thread only. However, the differences in implementation complexity are huge since the energy formulation is typically quite simple compared to the gradient and Hessian. 

\subsection{Comparison to classic automatic differentiation} 
\begin{figure}[t!]
	\begin{tikzpicture}
    \begin{axis}[
    xbar stacked,   
    xmin=0,         
    xmax=600,
    ymin=0.5,
    ymax=7.5,
    bar width = 8pt,
    width=1.1\linewidth,
    height=0.5\linewidth,
    ticks=none,
    legend columns = 2,
    legend style = {at={(0.5, -0.1)}, anchor=north, inner sep=3pt, style={column sep=0.15cm}},       
    legend cell align=left,  
    legend entries={{\lbrack Jakob 2010\rbrack},{\lbrack Desai et al. 2020\rbrack},{ours (basic)}, {ours (4 $\times$ simd)}, {ours (4 $\times$ simd, 8 threads)}, {ours (\device{GTX 1080})}, {ours (\device{Vega 20})}}
   ]

    \addplot[fill=bblue!40] coordinates {(508, 1)} node[right,pos=0.5]{508 ms};
    \resetstackedplots
    \addplot[fill=bblue!70] coordinates {(313, 2)} node[right,pos=0.5]{313 ms};
    \resetstackedplots
    \addplot[fill=rred!40] coordinates {(58.6, 3)} node[right,pos=0.5]{58.6 ms};
    \resetstackedplots
    \addplot[fill=rred!80] coordinates {(33, 4)} node[right,pos=0.5]{33 ms};
    \resetstackedplots
    \addplot[fill=rred!110] coordinates {(26.5, 5)} node[right,pos=0.5]{26.5 ms};
     \resetstackedplots
    \addplot[fill=ggreen!40] coordinates {(4.23, 6)} node[right,pos=0.5]{4.23 ms};
     \resetstackedplots
     \addplot[fill=ggreen] coordinates {(0.69, 7)} node[right,pos=0.5]{0.69 ms};
    
    \end{axis}
\end{tikzpicture}
	\caption{We compare to overloading based automatic differentiation \cite{Mitsuba} and a direct code transformation strategy \cite{Desai:2020}. All timings include the generation of Hessians and gradients per-face of a deformation energy for a mesh with 1M vertices. Due to our expression optimization techniques, we can outperform competing approaches even without enabling vectorization and parallelization. For the competing methods we run implementations provided by the authors of \cite{Desai:2020}. CPU timings taken on \device{Intel}.}
	\label{fig:adCompare}
\end{figure}
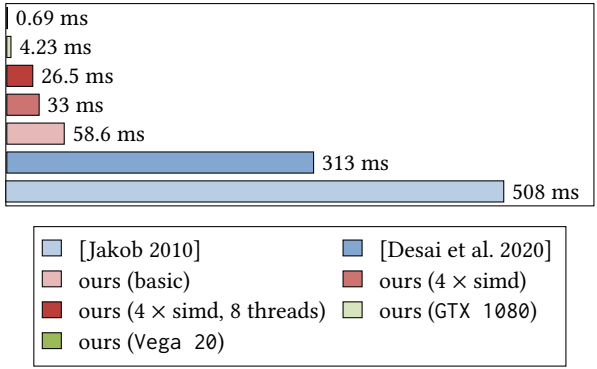
There exists a variety of automatic differentiation techniques. We compare here with implementations of the two most prevalent techniques, overloading based automatic differentiation \cite{Mitsuba}, also known as backpropagation, and direct source code transformation \cite{Desai:2020}. We compare using a benchmark example provided by Desai et al.~\shortcite{Desai:2020} that computes local Hessians of the symmetric Dirichlet deformation energy (\figref{fig:adCompare}). Our method outperforms both techniques. In contrast to \citet{Mitsuba} we do not have to maintain data structures to keep track of derivatives at runtime. Storing intermediate results and optimization expressions also gives us an advantage over direct code transformation techniques such as \citet{Desai:2020}.

\subsection{Comparison to linear algebra packages} \label{sec:compareLA}
Sparse matrix multiplication is supported by most linear algebra libraries and packages. We compare our method to competing methods on three platforms in Table \ref{tab:Lpow}. Highly optimized implementations of sparse matrix-matrix multiplication (spmm) can analyze a specific problem instance and try to find an optimal ordering and grouping of operations based on tiling strategies. These optimizations extensively use specific properties of matrix multiplication. Our method on the other hand is more general and is not aware of the fact that an spmm operation is computed. All methods that use problem specific optimizations in any form are marked with the $\dagger$ symbol (preprocessing is not included in the timings). All other methods use general purpose spmm methods. For MKL we include both variants: with preprocessing using the 2-stage matrix multiplication \texttt{mkl\_sparse\_sp2m} and without using \texttt{mkl\_sparse\_spmm}. Our method generates code that is faster than both versions. The taco compiler \cite{taco} can generate optimized and parallelized code from very concise descriptions of sparse linear algebra operations. For the comparison we used a hand optimized schedule with explicit storage of intermediate results as suggested by the authors \cite{kjolstad:2018}. Taco significantly improves over Eigen's and MKL's performance. On \device{Vega} we are able to outperform AMD's optimized sparse matrix algebra rocSparse \cite{rocmsparse}. However, on \device{GTX} cuSPARSE \cite{cusparse} is about 50\% faster than our method.

Apart from the timings we note that our method shares the advantage of taco that optimized code is automatically generated for different variants of matrices and operation flavors (transposing factors, computing only specific elements etc.). MKL, rocSparse and cuSparse support only some of these variants using many parameters. Performing spmm with cuSparse, for example, requires three function calls with about 10 parameters each and the allocation and management of several buffers \cite[15.6.16]{cusparse}.

\subsection{Ablation study} \label{sec:ablation}
\newcommand{\noimp}[1]{{\textcolor{lightgrey}{#1}}}
\begin{table*}[!t]
\begin{tabular}{@{}llllllllccl@{}}
\toprule
  &  & \multicolumn{1}{c}{+} & \multicolumn{1}{c}{+} & \multicolumn{1}{c}{+} & \multicolumn{1}{c}{+} & \multicolumn{1}{c}{+} & \multicolumn{1}{c}{+} & \multicolumn{1}{c}{+} & \multicolumn{1}{c}{+} &  \\
 & \multicolumn{1}{c}{\begin{tabular}[c]{@{}c@{}}reference\\ (Intel)\end{tabular}} & group & \multicolumn{1}{c}{\begin{tabular}[c]{@{}c@{}}explicit \\ tagging\end{tabular}} & \multicolumn{1}{c}{\begin{tabular}[c]{@{}c@{}}local\\ decompose\end{tabular}} & \multicolumn{1}{c}{\begin{tabular}[c]{@{}c@{}}global\\ decompose\end{tabular}} & \multicolumn{1}{c}{simplify} & \multicolumn{1}{c}{\begin{tabular}[c]{@{}c@{}}optimize\\ memory\end{tabular}} & \multicolumn{1}{c}{vectorize} & \multicolumn{1}{c}{\begin{tabular}[c]{@{}c@{}}parallelize\\ (8 threads)\end{tabular}} & \multicolumn{1}{c}{\begin{tabular}[c]{@{}c@{}}final\\ speedup\end{tabular}} \\ \hline

 \multicolumn{11}{l}{\textit{Cloth Hessian, 20k vertices, float, values only}} \\ \midrule
\device{Intel} & 136ms & 50ms & 13ms & 2.9ms & \noimp{2.9ms} & \noimp{2.9ms} & 2.7ms & \multicolumn{1}{l}{0.81ms} & \multicolumn{1}{l}{0.15ms} & 906x \\
Autodiff (\device{Intel}) & 136ms & \multicolumn{1}{c}{-} & 18ms & 4.9ms & \noimp{4.9ms} & 4.2ms & 3.9ms & \multicolumn{1}{l}{0.96ms} & \multicolumn{1}{l}{0.18ms} & 755x \\
\device{Vega} & 136ms & 28ms & 0.16ms & \noimp{0.16ms} & \noimp{0.16ms} & 0.15ms & 0.04ms & - & - & 3400x \\
Autodiff (\device{Vega}) & 136ms & \multicolumn{1}{c}{-} & 0.15ms & \noimp{0.15ms} & \noimp{0.15ms} & 0.14ms & 0.04ms & - & - & 3400x \\ \midrule
\multicolumn{11}{l}{\emph{Cloth Hessian, 20k vertices, float, values and sparse matrix construction}} \\ \midrule
\device{Intel} & 200ms & 50ms & 18ms & 7.5ms & \noimp{7.5ms} & \noimp{7.5ms} & 7.1ms & \multicolumn{1}{l}{5.8ms} & \multicolumn{1}{l}{1.4ms} & 142x \\
Autodiff (\device{Intel}) & 200ms & \multicolumn{1}{c}{-} & 24ms & 10ms & \noimp{10ms} & 9.0ms & \noimp{8.8ms} & \multicolumn{1}{l}{6.4ms} & \multicolumn{1}{l}{1.4ms} & 142x \\
\device{Vega} & 200ms & 27ms & 0.35ms & \noimp{0.35ms} & \noimp{0.35ms} & \noimp{0.35ms} & 0.24ms & - & - & 833x \\
Autodiff (\device{Vega}) & 200ms & \multicolumn{1}{c}{-} & 0.42ms & 0.35ms & \noimp{0.35ms} & \noimp{0.34ms} & 0.23ms & - & - & 870x \\ \midrule
\multicolumn{11}{l}{\textit{Cotan Laplacian, 200k vertices, float}} \\ \midrule
\device{Intel} & 105ms & 31ms & \multicolumn{1}{c}{-} & \noimp{31ms} & 15ms &     14ms & 13ms & \multicolumn{1}{l}{\noimp{13ms}} & \multicolumn{1}{l}{3.2ms} & 33x \\
\device{Vega} & 105ms & 1.62ms & \multicolumn{1}{c}{-} & 1.53 & 0.57 & \noimp{0.56} & 0.48 & - & - & 219x \\ \midrule
\multicolumn{11}{l}{$\mathbf L^3$, \emph{sparse}, $\mathbf L \in \mathds R^{50k\times 50k}$, \emph{float}} \\ \midrule
\device{Intel} & 145ms & 29ms & \multicolumn{1}{c}{-} & \noimp{28ms} & 12ms & \noimp{12ms} & \noimp{12ms} & \multicolumn{1}{l}{11ms} & \multicolumn{1}{l}{4.3ms} & 34x \\
\device{Vega} & 145ms & 1484ms & \multicolumn{1}{c}{-} & \noimp{1479ms} & 0.55ms & \noimp{0.55ms} & 0.49ms & - & - & 296x\\
\bottomrule
\end{tabular}
\caption{We performed an ablation study to assess the influence of our optimization steps. In each column we incrementally activate a new feature. We start with simple grouping, similar to EGGS \cite{eggs}. Next we allow for manually tagging \secref{sec:explicitIntermediates}, local (per kernel) and global expression decomposition to precompute redundant subexpressions \secref{sec:decomposition}, expression simplification \secref{sec:simplification}, memory optimizations as well as vectorization and parallelization \secref{sec:backend}. We report the overall speedup over the reference implementation measured on \device{Intel}. We report timings averaged over 100 iterations both for CPUs and a GPUs. For the computation of the stretch and shear energy from \cite{Baraff:1994} we compare hand written code and code generated by symbolic automatic differentiation. For timings that show less than 5\% performance improvement over the last optimization step we use the color gray. All experiments have been conducted on \device{Intel} with single precision. Speedups using double precision on \device{AMD} for the cotan and cloth Hessian example can be found in \figref{fig:cotan} and \figref{fig:clothAD}, respectively.}
\label{tab:ablation}
\end{table*}
The performance gains achieved by our method are a result of the combination of several optimization steps (\secref{sec:method}). We perform an ablation study to investigate how much effect the individual features have. To this end we activate them one by one until we arrive at the full method. The results are presented in Table~\ref{tab:ablation}. In all cases, the C++ compiler producing the final program is free to use all optimizations enabled with \texttt{-O3} and auto vectorization. Consequently all performance improvements are due to our optimization steps.
We show timings for the cloth example as well as for the construction of the cotan Laplacian and the computation of the third power of a sparse matrix. For the cloth energy Hessian we show two variants: using our method on hand optimized code and on expressions generated using symbolic automatic differentiation (\secref{sec:autodiff-comp}). The experiments are conducted for the computation of the per-element Hessians with and without sparse matrix construction. The reason for showing these experiments individually is their different performance behavior. The per-element Hessian computation is highly parallel and compute intensive while matrix assembly requires many memory accesses to sum up entries of the local Hessians.
In the following we elaborate on the effect individual optimization steps.

\paragraph{Grouping} The most basic approach is simple grouping. This means, we consider all output expressions as they are and sort them into structurally equivalent groups forming kernels. Unrolling these expressions already gives a considerable performance improvement; however, in many cases this produces an excessive amount of kernels that have a few instances only. For the last example, the computation of $\mathbf L^3$, basic grouping generates about 16 thousand kernels, most of them having one member only. 
The generated code files have a size of about 30 MB and need several hours to compile using clang. Since these kernels cannot easily be parallelized, they also do not profit from GPUs. 
\paragraph{Explicit tagging} We use explicit tagging only for the cloth examples, where we introduce a single function call for tagging the per-element $9 \times 9$ Hessian as intermediate (see \secref{sec:explicitIntermediates}). This ensures that they are all evaluated within one kernel making optimal use of local expression sharing. GPUs especially benefit from this step since the number of kernels and memory traffic are both greatly reduced.
\paragraph{Local decomposition} Using grouping with local decomposition roughly corresponds to the method implemented in EGGS \cite{eggs}. Caching common subexpression per kernel is especially useful for compute bound programs running on the CPU. While C++ compilers implement similar techniques, they are not as powerful as our algebraic decomposition method as witnessed by the speedup generated by this step.
\paragraph{Global decomposition} Being able to compute subexpressions such that they can be reused by other kernels not only saves time but also allows for a smaller number kernels in many cases. For the computation of $\mathbf L^3$, for example, the number of kernels is only 21 instead of 16000 without decomposition. Automatic decomposition also enables the efficient use of GPUs int this case. We do not see a benefit for computing the Hessian because explicit tagging of local Hessians already decomposes the computation in a nearly optimal way. For all other examples we see more than a factor of two in performance difference.
\paragraph{Simplification} The effect of expression simplification shows only in compute-bound scenarios. It is especially helpful for expressions generated by symbolic automatic differentiation. Other examples are based on code that has been optimized by hand and therefore does not result in expressions with a lot of optimization potential.
\paragraph{Memory optimization} Optimizing memory access encompasses several optimizations described in \secref{sec:backend}. One of the most effective allows for coalesced memory access when writing intermediate values or loading position indices and constants. The GPU version of the first example benefits the most from this method. 
The optimizations are not as efficient for the full Hessian matrix construction because most memory accesses load intermediates that cannot be optimized. The reason is that the storage locations for intermediates have already been fixed by our system when writing their values.
\paragraph{Vectorization} We support vectorization by using nested loops and \emph{blocked} coalesced memory access as described in \secref{sec:backend}. Again, compute-bound algorithms benefit the most from this feature. Note that auto-vectorization of the C++ compiler is activated for all tests but really makes a difference when used together with our vectorization-friendly code and memory optimizations. 
\paragraph{Parallelization} Since all instances of a kernel can be evaluated independently, we can trivially employ parallelization and obtain considerable speedups. This also helps memory-bound application by using multiple memory controllers.

\vspace{12pt}
\noindent The ablation study shows that the performance boosts due to different optimization steps vary depending on the device and type of algorithm. However, no optimization step decreases performance. Combining all stages makes for a versatile code optimization technique that tries to use features of the target device as efficiently as possible.

\section{Limitations and Concluding Remarks}
We introduced a compiler for algorithms with a static execution tree that shows that speedups of two to three orders of magnitude can be achieved by unrolling the computational tree, and then regrouping expressions in a hardware friendly way.
Our reference implementation shows that this approach is very promising and practical. There are, however, a few engineering aspects that would require additional work to make this method more accessible.
(1) The numerical types of the code need to be templates to be compatible with our symbolic type. Automatically replacing types for non-templated code would make our method even more practical.
(2) The central limitation of our method is that it is time and memory expensive. Therefore it is useful if the code to be optimized is executed many times. For this paper our focus was on reducing the running time, however, code optimization time as well as memory consumption can likely be reduced by a significant factor by optimizing and parallelizing the pipeline and compressing expression trees. (3) The current compilation process is serial and not optimized. We believe that with more engineering the compilation time and memory requirements could be dramatically reduced. 
(4) Finally we would like to remove the need to automatically annotate dense blocks in order to evaluate several expressions in a single kernel. This could be achieved by searching all expression trees for potential dense blocks.

We believe that the main avenue for future work, apart from the points listed below, is to extend our approach to target multiple processors and accelerators, while taking into account the latency in the channels connecting them. The idea is to add to the input a description of the system (for example, a pair of GPUs in the same machine connected via PCIExpress, or a collection of 10 workstations in a local ethernet network), and generate an optimal execution policy that distributes the data and computation across all the available resources by taking advantage of the knowledge of the entire computational graph. We believe this is an exciting direction that could dramatically lower the barrier for developing research code on heterogeneous HPC clusters using a combination of CPUs, GPUs, and TPUs.

\subsection{Compiler pass vs. code generator}
A key design decision for our tool was to build a code generator leveraging phased compilation, rather than integrating our methodology into an existing compiler infrastructure such as LLVM; for example, we could have used an approach similar to profile-guided optimization to first compile code with instrumentation that generates traces for a second compilation step. Integrating within LLVM would have some advantages, including implementations of several optimization techniques we utilize.    Instead of generating C++ source code, such an approach could generate LLVM IR.

Alternatively, we could have attempted a purely-static approach that, instead of utilizing a two-phase compilation approach, used the combination of input values and LLVM IR to interpret or otherwise symbolically determine expression trees based on the input values at compile-time.  With this approach, we would need to either utilize LLVM’s own interpretation framework or write our own, along with infrastructure to read sparse values from input files.

In either alternative design, much of the the framework would be similar to our code generator: we still would need to process expression trees or similar data structures, and perform expression grouping and leaf harvesting. Expression optimization could utilize existing pieces of the LLVM framework, but we would likely need to write our own passes as well.  In addition, the alternate designs would require generating LLVM IR rather than strings.  In the end, we rejected these alternatives due to the simple fact that they would require writing more code than for our approach. We leave integrating with LLVM or another compiler infrastructure, which could lead to additional speedups and wider applicability, to future work.

\section*{Acknowledgements}
We thank the NYU IT High Performance Computing for resources, services, and staff expertise. This work was partially supported by the NSF CAREER award under Grant No. 1652515, the NSF grants OAC-1835712, OIA-1937043, CHS-1908767, CHS-1901091, NSERC DGECR-2021-00461 and RGPIN-2021-03707, a Sloan Fellowship, a gift from Adobe Research, a gift from Advanced Micro Devices, Inc and by the European Research Council (ERC) under the European Union’s Horizon 2020 research and innovation programme (grant agreement No. 101003104).

\bibliographystyle{ACM-Reference-Format}
\bibliography{main}


\begin{thebibliography}{53}


\ifx \showCODEN    \undefined \def \showCODEN     #1{\unskip}     \fi
\ifx \showDOI      \undefined \def \showDOI       #1{#1}\fi
\ifx \showISBNx    \undefined \def \showISBNx     #1{\unskip}     \fi
\ifx \showISBNxiii \undefined \def \showISBNxiii  #1{\unskip}     \fi
\ifx \showISSN     \undefined \def \showISSN      #1{\unskip}     \fi
\ifx \showLCCN     \undefined \def \showLCCN      #1{\unskip}     \fi
\ifx \shownote     \undefined \def \shownote      #1{#1}          \fi
\ifx \showarticletitle \undefined \def \showarticletitle #1{#1}   \fi
\ifx \showURL      \undefined \def \showURL       {\relax}        \fi
\providecommand\bibfield[2]{#2}
\providecommand\bibinfo[2]{#2}
\providecommand\natexlab[1]{#1}
\providecommand\showeprint[2][]{arXiv:#2}

\bibitem[\protect\citeauthoryear{Aho, Lam, Sethi, and Ullman}{Aho
  et~al\mbox{.}}{2013}]%
        {aho2013}
\bibfield{author}{\bibinfo{person}{AV Aho}, \bibinfo{person}{Monica~S Lam},
  \bibinfo{person}{R Sethi}, {and} \bibinfo{person}{JD Ullman}.}
  \bibinfo{year}{2013}\natexlab{}.
\newblock \bibinfo{booktitle}{\emph{Compilers: Pearson New International
  Edition: Principles, Techniques, and Tools}}.
\newblock \bibinfo{publisher}{Pearson}.
\newblock


\bibitem[\protect\citeauthoryear{Alexa, Herholz, Kohlbrenner, and
  Sorkine-Hornung}{Alexa et~al\mbox{.}}{2020}]%
        {Alexa:PLOTM:2020}
\bibfield{author}{\bibinfo{person}{Marc Alexa}, \bibinfo{person}{Philipp
  Herholz}, \bibinfo{person}{Maximilian Kohlbrenner}, {and}
  \bibinfo{person}{Olga Sorkine-Hornung}.} \bibinfo{year}{2020}\natexlab{}.
\newblock \showarticletitle{Properties of {L}aplace Operators for Tetrahedral
  Meshes}.
\newblock \bibinfo{journal}{\emph{Computer Graphics Forum (proceedings of SGP
  2020)}} \bibinfo{volume}{39}, \bibinfo{number}{5} (\bibinfo{year}{2020}),
  \bibinfo{numpages}{12}~pages.
\newblock
\showISSN{1467-8659}
\urldef\tempurl%
\url{https://doi.org/10.1111/cgf.14068}
\showDOI{\tempurl}


\bibitem[\protect\citeauthoryear{{AMD}}{{AMD}}{2020}]%
        {rocmsparse}
\bibfield{author}{\bibinfo{person}{{AMD}}.} \bibinfo{year}{2020}\natexlab{}.
\newblock \bibinfo{title}{rocmSPARSE}.
\newblock
\newblock
\urldef\tempurl%
\url{https://rocsparse.readthedocs.io/en/master/}
\showURL{%
\tempurl}


\bibitem[\protect\citeauthoryear{Anderson, Bai, Bischof, Blackford, Demmel,
  Dongarra, Du~Croz, Greenbaum, Hammarling, McKenney, and Sorensen}{Anderson
  et~al\mbox{.}}{1999}]%
        {LAPACK}
\bibfield{author}{\bibinfo{person}{E. Anderson}, \bibinfo{person}{Z. Bai},
  \bibinfo{person}{C. Bischof}, \bibinfo{person}{S. Blackford},
  \bibinfo{person}{J. Demmel}, \bibinfo{person}{J. Dongarra},
  \bibinfo{person}{J. Du~Croz}, \bibinfo{person}{A. Greenbaum},
  \bibinfo{person}{S. Hammarling}, \bibinfo{person}{A. McKenney}, {and}
  \bibinfo{person}{D. Sorensen}.} \bibinfo{year}{1999}\natexlab{}.
\newblock \bibinfo{booktitle}{\emph{{LAPACK} Users' Guide}
  (\bibinfo{edition}{third} ed.)}.
\newblock \bibinfo{publisher}{Society for Industrial and Applied Mathematics},
  \bibinfo{address}{Philadelphia, PA}.
\newblock
\showISBNx{0-89871-447-8 (paperback)}


\bibitem[\protect\citeauthoryear{Augustine, Sarma, Pouchet, and
  Rodr\'{\i}guez}{Augustine et~al\mbox{.}}{2019}]%
        {augustinepldi2019}
\bibfield{author}{\bibinfo{person}{Travis Augustine},
  \bibinfo{person}{Janarthanan Sarma}, \bibinfo{person}{Louis-No\"{e}l
  Pouchet}, {and} \bibinfo{person}{Gabriel Rodr\'{\i}guez}.}
  \bibinfo{year}{2019}\natexlab{}.
\newblock \showarticletitle{Generating Piecewise-Regular Code from Irregular
  Structures}. In \bibinfo{booktitle}{\emph{Proceedings of the 40th ACM SIGPLAN
  Conference on Programming Language Design and Implementation}} (Phoenix, AZ,
  USA) \emph{(\bibinfo{series}{PLDI 2019})}. \bibinfo{publisher}{Association
  for Computing Machinery}, \bibinfo{address}{New York, NY, USA},
  \bibinfo{pages}{625–639}.
\newblock
\showISBNx{9781450367127}
\urldef\tempurl%
\url{https://doi.org/10.1145/3314221.3314615}
\showDOI{\tempurl}


\bibitem[\protect\citeauthoryear{Balay, Gropp, McInnes, and Smith}{Balay
  et~al\mbox{.}}{1997}]%
        {petsc}
\bibfield{author}{\bibinfo{person}{Satish Balay}, \bibinfo{person}{William~D
  Gropp}, \bibinfo{person}{Lois~Curfman McInnes}, {and}
  \bibinfo{person}{Barry~F Smith}.} \bibinfo{year}{1997}\natexlab{}.
\newblock \showarticletitle{Efficient management of parallelism in
  object-oriented numerical software libraries}.
\newblock In \bibinfo{booktitle}{\emph{Modern software tools for scientific
  computing}}. \bibinfo{publisher}{Springer}, \bibinfo{address}{Birkh{\"a}user
  Boston}, \bibinfo{pages}{163--202}.
\newblock


\bibitem[\protect\citeauthoryear{Baraff and Witkin}{Baraff and Witkin}{1998}]%
        {Baraff:1994}
\bibfield{author}{\bibinfo{person}{David Baraff} {and} \bibinfo{person}{Andrew
  Witkin}.} \bibinfo{year}{1998}\natexlab{}.
\newblock \showarticletitle{Large Steps in Cloth Simulation}. In
  \bibinfo{booktitle}{\emph{Proceedings of the 25th Annual Conference on
  Computer Graphics and Interactive Techniques}}
  \emph{(\bibinfo{series}{SIGGRAPH '98})}. \bibinfo{publisher}{Association for
  Computing Machinery}, \bibinfo{address}{New York, NY, USA},
  \bibinfo{pages}{43–54}.
\newblock
\showISBNx{0897919998}
\urldef\tempurl%
\url{https://doi.org/10.1145/280814.280821}
\showDOI{\tempurl}


\bibitem[\protect\citeauthoryear{Baydin, Pearlmutter, Radul, and
  Siskind}{Baydin et~al\mbox{.}}{2018}]%
        {baydin2018automatic}
\bibfield{author}{\bibinfo{person}{Atilim~Gunes Baydin},
  \bibinfo{person}{Barak~A Pearlmutter}, \bibinfo{person}{Alexey~Andreyevich
  Radul}, {and} \bibinfo{person}{Jeffrey~Mark Siskind}.}
  \bibinfo{year}{2018}\natexlab{}.
\newblock \showarticletitle{Automatic differentiation in machine learning: a
  survey}.
\newblock \bibinfo{journal}{\emph{Journal of machine learning research}}
  \bibinfo{volume}{18} (\bibinfo{year}{2018}).
\newblock


\bibitem[\protect\citeauthoryear{Bernstein, Shah, Lemire, Devito, Fisher,
  Levis, and Hanrahan}{Bernstein et~al\mbox{.}}{2016}]%
        {ebb}
\bibfield{author}{\bibinfo{person}{Gilbert~Louis Bernstein},
  \bibinfo{person}{Chinmayee Shah}, \bibinfo{person}{Crystal Lemire},
  \bibinfo{person}{Zachary Devito}, \bibinfo{person}{Matthew Fisher},
  \bibinfo{person}{Philip Levis}, {and} \bibinfo{person}{Pat Hanrahan}.}
  \bibinfo{year}{2016}\natexlab{}.
\newblock \showarticletitle{Ebb: A DSL for Physical Simulation on CPUs and
  GPUs}.
\newblock  \bibinfo{volume}{35}, \bibinfo{number}{2}, Article
  \bibinfo{articleno}{21} (\bibinfo{date}{may} \bibinfo{year}{2016}),
  \bibinfo{numpages}{12}~pages.
\newblock
\showISSN{0730-0301}
\urldef\tempurl%
\url{https://doi.org/10.1145/2892632}
\showDOI{\tempurl}


\bibitem[\protect\citeauthoryear{Bezanson, Karpinski, Shah, and
  Edelman}{Bezanson et~al\mbox{.}}{2012}]%
        {julia}
\bibfield{author}{\bibinfo{person}{Jeff Bezanson}, \bibinfo{person}{Stefan
  Karpinski}, \bibinfo{person}{Viral~B. Shah}, {and} \bibinfo{person}{Alan
  Edelman}.} \bibinfo{year}{2012}\natexlab{}.
\newblock \bibinfo{title}{{J}ulia: {A} Fast Dynamic Language for Technical
  Computing}.
\newblock
\newblock


\bibitem[\protect\citeauthoryear{Bik and Wijshoff}{Bik and Wijshoff}{1993}]%
        {dutch1}
\bibfield{author}{\bibinfo{person}{Aart~JC Bik} {and} \bibinfo{person}{Harry~AG
  Wijshoff}.} \bibinfo{year}{1993}\natexlab{}.
\newblock \showarticletitle{Compilation techniques for sparse matrix
  computations}. In \bibinfo{booktitle}{\emph{Proceedings of the 7th
  international conference on Supercomputing}}. ACM, \bibinfo{pages}{416--424}.
\newblock


\bibitem[\protect\citeauthoryear{Bik and Wijshoff}{Bik and Wijshoff}{1994}]%
        {dutch2}
\bibfield{author}{\bibinfo{person}{Aart~JC Bik} {and} \bibinfo{person}{Harry~AG
  Wijshoff}.} \bibinfo{year}{1994}\natexlab{}.
\newblock \showarticletitle{On automatic data structure selection and code
  generation for sparse computations}.
\newblock In \bibinfo{booktitle}{\emph{Languages and Compilers for Parallel
  Computing}}. \bibinfo{publisher}{Springer}, \bibinfo{pages}{57--75}.
\newblock


\bibitem[\protect\citeauthoryear{Bollh{\"o}fer, Schenk, Janalik, Hamm, and
  Gullapalli}{Bollh{\"o}fer et~al\mbox{.}}{2020}]%
        {pardiso}
\bibfield{author}{\bibinfo{person}{Matthias Bollh{\"o}fer},
  \bibinfo{person}{Olaf Schenk}, \bibinfo{person}{Radim Janalik},
  \bibinfo{person}{Steve Hamm}, {and} \bibinfo{person}{Kiran Gullapalli}.}
  \bibinfo{year}{2020}\natexlab{}.
\newblock \showarticletitle{State-of-the-Art Sparse Direct Solvers}.
\newblock  (\bibinfo{year}{2020}), \bibinfo{pages}{3--33}.
\newblock
\showISBNx{978-3-030-43736-7}
\urldef\tempurl%
\url{https://doi.org/10.1007/978-3-030-43736-7_1}
\showDOI{\tempurl}


\bibitem[\protect\citeauthoryear{Byun, Lin, Yelick, and Demmel}{Byun
  et~al\mbox{.}}{2012}]%
        {byun2012}
\bibfield{author}{\bibinfo{person}{Jong-Ho Byun}, \bibinfo{person}{Richard
  Lin}, \bibinfo{person}{Katherine~A Yelick}, {and} \bibinfo{person}{James
  Demmel}.} \bibinfo{year}{2012}\natexlab{}.
\newblock \showarticletitle{Autotuning sparse matrix-vector multiplication for
  multicore}.
\newblock \bibinfo{journal}{\emph{EECS, UC Berkeley, Tech. Rep}}
  (\bibinfo{year}{2012}).
\newblock


\bibitem[\protect\citeauthoryear{Cheshmi, Kamil, Strout, and Dehnavi}{Cheshmi
  et~al\mbox{.}}{2018}]%
        {cheshmi}
\bibfield{author}{\bibinfo{person}{Kazem Cheshmi}, \bibinfo{person}{Shoaib
  Kamil}, \bibinfo{person}{Michelle~Mills Strout}, {and}
  \bibinfo{person}{Maryam~Mehri Dehnavi}.} \bibinfo{year}{2018}\natexlab{}.
\newblock \showarticletitle{ParSy: Inspection and Transformation of Sparse
  Matrix Computations for Parallelism}. In
  \bibinfo{booktitle}{\emph{Proceedings of the International Conference for
  High Performance Computing, Networking, Storage, and Analysis}} (Dallas,
  Texas) \emph{(\bibinfo{series}{SC '18})}. \bibinfo{publisher}{IEEE Press},
  \bibinfo{address}{Piscataway, NJ, USA}, Article \bibinfo{articleno}{62},
  \bibinfo{numpages}{15}~pages.
\newblock
\urldef\tempurl%
\url{http://dl.acm.org/citation.cfm?id=3291656.3291739}
\showURL{%
\tempurl}


\bibitem[\protect\citeauthoryear{Chou, Kjolstad, and Amarasinghe}{Chou
  et~al\mbox{.}}{2018}]%
        {taco-stephen}
\bibfield{author}{\bibinfo{person}{Stephen Chou}, \bibinfo{person}{Fredrik
  Kjolstad}, {and} \bibinfo{person}{Saman Amarasinghe}.}
  \bibinfo{year}{2018}\natexlab{}.
\newblock \showarticletitle{Format Abstraction for Sparse Tensor Algebra
  Compilers}.
\newblock \bibinfo{journal}{\emph{Proc. ACM Program. Lang.}}
  \bibinfo{volume}{2}, \bibinfo{number}{OOPSLA}, Article
  \bibinfo{articleno}{123} (\bibinfo{date}{Oct.} \bibinfo{year}{2018}),
  \bibinfo{numpages}{30}~pages.
\newblock
\showISSN{2475-1421}
\urldef\tempurl%
\url{https://doi.org/10.1145/3276493}
\showDOI{\tempurl}


\bibitem[\protect\citeauthoryear{Desai, Shuchatowitz, Jiang, Schneider, and
  Panozzo}{Desai et~al\mbox{.}}{2020}]%
        {Desai:2020}
\bibfield{author}{\bibinfo{person}{Deshana Desai}, \bibinfo{person}{Etai
  Shuchatowitz}, \bibinfo{person}{Zhongshi Jiang}, \bibinfo{person}{Teseo
  Schneider}, {and} \bibinfo{person}{Daniele Panozzo}.}
  \bibinfo{year}{2020}\natexlab{}.
\newblock \bibinfo{title}{ACORNS: An Easy-To-Use Code Generator for Gradients
  and Hessians}.
\newblock
\newblock
\showeprint[arxiv]{2007.05094}~[cs.MS]


\bibitem[\protect\citeauthoryear{Devito, Mara, Zollh\"{o}fer, Bernstein,
  Ragan-Kelley, Theobalt, Hanrahan, Fisher, and Niessner}{Devito
  et~al\mbox{.}}{2017}]%
        {opt}
\bibfield{author}{\bibinfo{person}{Zachary Devito}, \bibinfo{person}{Michael
  Mara}, \bibinfo{person}{Michael Zollh\"{o}fer}, \bibinfo{person}{Gilbert
  Bernstein}, \bibinfo{person}{Jonathan Ragan-Kelley},
  \bibinfo{person}{Christian Theobalt}, \bibinfo{person}{Pat Hanrahan},
  \bibinfo{person}{Matthew Fisher}, {and} \bibinfo{person}{Matthias Niessner}.}
  \bibinfo{year}{2017}\natexlab{}.
\newblock \showarticletitle{Opt: A Domain Specific Language for Non-Linear
  Least Squares Optimization in Graphics and Imaging}.
\newblock \bibinfo{journal}{\emph{ACM Trans. Graph.}} \bibinfo{volume}{36},
  \bibinfo{number}{5}, Article \bibinfo{articleno}{171} (\bibinfo{date}{oct}
  \bibinfo{year}{2017}), \bibinfo{numpages}{27}~pages.
\newblock
\showISSN{0730-0301}
\urldef\tempurl%
\url{https://doi.org/10.1145/3132188}
\showDOI{\tempurl}


\bibitem[\protect\citeauthoryear{Feautrier}{Feautrier}{1988}]%
        {FEAUTRIER1988ARRAY}
\bibfield{author}{\bibinfo{person}{Paul Feautrier}.}
  \bibinfo{year}{1988}\natexlab{}.
\newblock \showarticletitle{Array Expansion}. In \bibinfo{booktitle}{\emph{2nd
  International Conference on Supercomputing (ICS'88)}}. ACM,
  \bibinfo{pages}{429--441}.
\newblock


\bibitem[\protect\citeauthoryear{Google}{Google}{2017}]%
        {tensorflow-sparse}
\bibfield{author}{\bibinfo{person}{Google}.} \bibinfo{year}{2017}\natexlab{}.
\newblock \bibinfo{title}{TensorFlow Sparse Tensors}.
\newblock
  \bibinfo{howpublished}{\url{https://www.tensorflow.org/api_guides/python/sparse_ops}}.
\newblock


\bibitem[\protect\citeauthoryear{Goto and {v}an~{d}e Geijn}{Goto and {v}an~{d}e
  Geijn}{2008}]%
        {goto}
\bibfield{author}{\bibinfo{person}{Kazushige Goto} {and}
  \bibinfo{person}{Robert~A. {v}an~{d}e Geijn}.}
  \bibinfo{year}{2008}\natexlab{}.
\newblock \showarticletitle{Anatomy of High-Performance Matrix Multiplication}.
\newblock \bibinfo{journal}{\emph{ACM Trans. Math. Softw.}}
  \bibinfo{volume}{34}, \bibinfo{number}{3}, Article \bibinfo{articleno}{12}
  (\bibinfo{date}{May} \bibinfo{year}{2008}), \bibinfo{numpages}{25}~pages.
\newblock
\showISSN{0098-3500}


\bibitem[\protect\citeauthoryear{Griewank and Walther}{Griewank and
  Walther}{2008}]%
        {griewank2008evaluating}
\bibfield{author}{\bibinfo{person}{Andreas Griewank} {and}
  \bibinfo{person}{Andrea Walther}.} \bibinfo{year}{2008}\natexlab{}.
\newblock \bibinfo{booktitle}{\emph{Evaluating derivatives: principles and
  techniques of algorithmic differentiation}}. Vol.~\bibinfo{volume}{105}.
\newblock \bibinfo{publisher}{Siam}.
\newblock


\bibitem[\protect\citeauthoryear{Guennebaud, Jacob, et~al\mbox{.}}{Guennebaud
  et~al\mbox{.}}{2010}]%
        {eigen}
\bibfield{author}{\bibinfo{person}{Ga\"{e}l Guennebaud},
  \bibinfo{person}{Beno\^{i}t Jacob}, {et~al\mbox{.}}}
  \bibinfo{year}{2010}\natexlab{}.
\newblock \bibinfo{title}{Eigen v3}.
\newblock \bibinfo{howpublished}{\url{http://eigen.tuxfamily.org}}.
\newblock


\bibitem[\protect\citeauthoryear{Iglberger, Hager, Treibig, and
  Rüde}{Iglberger et~al\mbox{.}}{2012}]%
        {iglberger2012}
\bibfield{author}{\bibinfo{person}{Klaus Iglberger}, \bibinfo{person}{Georg
  Hager}, \bibinfo{person}{Jan Treibig}, {and} \bibinfo{person}{Ulrich Rüde}.}
  \bibinfo{year}{2012}\natexlab{}.
\newblock \showarticletitle{High performance smart expression template math
  libraries}. In \bibinfo{booktitle}{\emph{2012 International Conference on
  High Performance Computing Simulation (HPCS)}}. \bibinfo{pages}{367--373}.
\newblock
\urldef\tempurl%
\url{https://doi.org/10.1109/HPCSim.2012.6266939}
\showDOI{\tempurl}


\bibitem[\protect\citeauthoryear{Intel}{Intel}{2021}]%
        {mkl}
\bibfield{author}{\bibinfo{person}{Intel}.} \bibinfo{year}{2021}\natexlab{}.
\newblock \bibinfo{booktitle}{\emph{Developer reference for Intel oneAPI Math
  Kernel Library - C}}.
\newblock \bibinfo{type}{{T}echnical {R}eport}.
  \bibinfo{institution}{\url{https://software.intel.com/content/dam/develop/external/us/en/documents/onemkl-developerreference-c.pdf}}.
\newblock


\bibitem[\protect\citeauthoryear{Jakob}{Jakob}{2010}]%
        {Mitsuba}
\bibfield{author}{\bibinfo{person}{Wenzel Jakob}.}
  \bibinfo{year}{2010}\natexlab{}.
\newblock \bibinfo{title}{Mitsuba renderer}.
\newblock
\newblock
\newblock
\shownote{http://www.mitsuba-renderer.org.}


\bibitem[\protect\citeauthoryear{Kjolstad, Ahrens, Kamil, and
  Amarasinghe}{Kjolstad et~al\mbox{.}}{2019}]%
        {kjolstad:2018}
\bibfield{author}{\bibinfo{person}{Fredrik Kjolstad}, \bibinfo{person}{Peter
  Ahrens}, \bibinfo{person}{Shoaib Kamil}, {and} \bibinfo{person}{Saman
  Amarasinghe}.} \bibinfo{year}{2019}\natexlab{}.
\newblock \showarticletitle{Tensor Algebra Compilation with Workspaces}.
\newblock  (\bibinfo{year}{2019}), \bibinfo{pages}{180--192}.
\newblock
\showISBNx{978-1-7281-1436-1}
\urldef\tempurl%
\url{http://dl.acm.org/citation.cfm?id=3314872.3314894}
\showURL{%
\tempurl}


\bibitem[\protect\citeauthoryear{Kjolstad, Kamil, Chou, Lugato, and
  Amarasinghe}{Kjolstad et~al\mbox{.}}{2017}]%
        {taco}
\bibfield{author}{\bibinfo{person}{Fredrik Kjolstad}, \bibinfo{person}{Shoaib
  Kamil}, \bibinfo{person}{Stephen Chou}, \bibinfo{person}{David Lugato}, {and}
  \bibinfo{person}{Saman Amarasinghe}.} \bibinfo{year}{2017}\natexlab{}.
\newblock \showarticletitle{The Tensor Algebra Compiler}.
\newblock \bibinfo{journal}{\emph{Proc. ACM Program. Lang.}}
  \bibinfo{volume}{1}, \bibinfo{number}{OOPSLA}, Article
  \bibinfo{articleno}{77} (\bibinfo{date}{Oct.} \bibinfo{year}{2017}),
  \bibinfo{numpages}{29}~pages.
\newblock
\showISSN{2475-1421}
\urldef\tempurl%
\url{https://doi.org/10.1145/3133901}
\showDOI{\tempurl}


\bibitem[\protect\citeauthoryear{Kjolstad, Kamil, Ragan-Kelley, Levin, Sueda,
  Chen, Vouga, Kaufman, Kanwar, Matusik, and Amarasinghe}{Kjolstad
  et~al\mbox{.}}{2016}]%
        {simit}
\bibfield{author}{\bibinfo{person}{Fredrik Kjolstad}, \bibinfo{person}{Shoaib
  Kamil}, \bibinfo{person}{Jonathan Ragan-Kelley}, \bibinfo{person}{David
  Levin}, \bibinfo{person}{Shinjiro Sueda}, \bibinfo{person}{Desai Chen},
  \bibinfo{person}{Etienne Vouga}, \bibinfo{person}{Danny Kaufman},
  \bibinfo{person}{Gurtej Kanwar}, \bibinfo{person}{Wojciech Matusik}, {and}
  \bibinfo{person}{Saman Amarasinghe}.} \bibinfo{year}{2016}\natexlab{}.
\newblock \showarticletitle{Simit: A Language for Physical Simulation}.
\newblock \bibinfo{journal}{\emph{ACM Trans. Graphics}} (\bibinfo{year}{2016}).
\newblock


\bibitem[\protect\citeauthoryear{Kotlyar, Pingali, and Stodghill}{Kotlyar
  et~al\mbox{.}}{1997}]%
        {bernoulli}
\bibfield{author}{\bibinfo{person}{Vladimir Kotlyar}, \bibinfo{person}{Keshav
  Pingali}, {and} \bibinfo{person}{Paul Stodghill}.}
  \bibinfo{year}{1997}\natexlab{}.
\newblock \showarticletitle{A relational approach to the compilation of sparse
  matrix programs}.
\newblock In \bibinfo{booktitle}{\emph{Euro-Par'97 Parallel Processing}}.
  \bibinfo{publisher}{Springer}, \bibinfo{pages}{318--327}.
\newblock


\bibitem[\protect\citeauthoryear{Liu, Wang, Foroosh, Tappen, and Pensky}{Liu
  et~al\mbox{.}}{2015}]%
        {Liu_2015_CVPR}
\bibfield{author}{\bibinfo{person}{Baoyuan Liu}, \bibinfo{person}{Min Wang},
  \bibinfo{person}{Hassan Foroosh}, \bibinfo{person}{Marshall Tappen}, {and}
  \bibinfo{person}{Marianna Pensky}.} \bibinfo{year}{2015}\natexlab{}.
\newblock \showarticletitle{Sparse Convolutional Neural Networks}. In
  \bibinfo{booktitle}{\emph{Proceedings of the IEEE Conference on Computer
  Vision and Pattern Recognition (CVPR)}}.
\newblock


\bibitem[\protect\citeauthoryear{Mahmoud, Porumbescu, and Owens}{Mahmoud
  et~al\mbox{.}}{2021}]%
        {rxmesh}
\bibfield{author}{\bibinfo{person}{Ahmed~H. Mahmoud},
  \bibinfo{person}{Serban~D. Porumbescu}, {and} \bibinfo{person}{John~D.
  Owens}.} \bibinfo{year}{2021}\natexlab{}.
\newblock \showarticletitle{{RXM}esh: A {GPU} Mesh Data Structure}.
\newblock \bibinfo{journal}{\emph{ACM Transactions on Graphics}}
  \bibinfo{volume}{40}, \bibinfo{number}{4} (\bibinfo{date}{Aug.}
  \bibinfo{year}{2021}).
\newblock
\urldef\tempurl%
\url{https://doi.org/10.1145/3450626.3459748}
\showDOI{\tempurl}


\bibitem[\protect\citeauthoryear{Mara, Heide, Zollh\"{o}fer, Nie\ss{}ner, and
  Hanrahan}{Mara et~al\mbox{.}}{2021}]%
        {thallo}
\bibfield{author}{\bibinfo{person}{Michael Mara}, \bibinfo{person}{Felix
  Heide}, \bibinfo{person}{Michael Zollh\"{o}fer}, \bibinfo{person}{Matthias
  Nie\ss{}ner}, {and} \bibinfo{person}{Pat Hanrahan}.}
  \bibinfo{year}{2021}\natexlab{}.
\newblock \showarticletitle{Thallo – Scheduling for High-Performance
  Large-Scale Non-Linear Least-Squares Solvers}.
\newblock \bibinfo{journal}{\emph{ACM Trans. Graph.}} \bibinfo{volume}{40},
  \bibinfo{number}{5}, Article \bibinfo{articleno}{184} (\bibinfo{date}{sep}
  \bibinfo{year}{2021}), \bibinfo{numpages}{14}~pages.
\newblock
\showISSN{0730-0301}
\urldef\tempurl%
\url{https://doi.org/10.1145/3453986}
\showDOI{\tempurl}


\bibitem[\protect\citeauthoryear{MATLAB}{MATLAB}{2014}]%
        {matlab}
\bibfield{author}{\bibinfo{person}{MATLAB}.} \bibinfo{year}{2014}\natexlab{}.
\newblock \bibinfo{booktitle}{\emph{version 8.3.0 (R2014a)}}.
\newblock \bibinfo{publisher}{The MathWorks Inc.}, \bibinfo{address}{Natick,
  Massachusetts}.
\newblock


\bibitem[\protect\citeauthoryear{McAdams, Selle, Tamstorf, Teran, and
  Sifakis}{McAdams et~al\mbox{.}}{2011}]%
        {McAdams:2011}
\bibfield{author}{\bibinfo{person}{Aleka McAdams}, \bibinfo{person}{Andrew
  Selle}, \bibinfo{person}{Rasmus Tamstorf}, \bibinfo{person}{Joseph Teran},
  {and} \bibinfo{person}{Eftychios Sifakis}.} \bibinfo{year}{2011}\natexlab{}.
\newblock \bibinfo{booktitle}{\emph{Computing the singular value decomposition
  of 3x3 matrices with minimal branching and elementary floating point
  operations}}.
\newblock \bibinfo{type}{{T}echnical {R}eport}.
  \bibinfo{institution}{University of Wisconsin-Madison Department of Computer
  Sciences}.
\newblock


\bibitem[\protect\citeauthoryear{Nvidia}{Nvidia}{2020}]%
        {cusparse}
\bibfield{author}{\bibinfo{person}{Nvidia}.} \bibinfo{year}{2020}\natexlab{}.
\newblock \bibinfo{booktitle}{\emph{cuSPARSE Library}}.
\newblock \bibinfo{type}{{T}echnical {R}eport}.
  \bibinfo{institution}{\url{https://docs.nvidia.com/cuda/pdf/CUSPARSE_Library.pdf}}.
\newblock


\bibitem[\protect\citeauthoryear{Pugh}{Pugh}{1991}]%
        {PUGH1991UNIFORM}
\bibfield{author}{\bibinfo{person}{William Pugh}.}
  \bibinfo{year}{1991}\natexlab{}.
\newblock \showarticletitle{Uniform Techniques for Loop Optimization}. In
  \bibinfo{booktitle}{\emph{5th International Conference on Supercomputing
  (ICS'91)}}. ACM, \bibinfo{pages}{341--352}.
\newblock


\bibitem[\protect\citeauthoryear{Ragan-Kelley, Adams, Paris, Levoy,
  Amarasinghe, and Durand}{Ragan-Kelley et~al\mbox{.}}{2012}]%
        {halide-siggraph}
\bibfield{author}{\bibinfo{person}{Jonathan Ragan-Kelley},
  \bibinfo{person}{Andrew Adams}, \bibinfo{person}{Sylvain Paris},
  \bibinfo{person}{Marc Levoy}, \bibinfo{person}{Saman Amarasinghe}, {and}
  \bibinfo{person}{Fr\'{e}do Durand}.} \bibinfo{year}{2012}\natexlab{}.
\newblock \showarticletitle{Decoupling Algorithms from Schedules for Easy
  Optimization of Image Processing Pipelines}.
\newblock \bibinfo{journal}{\emph{ACM Trans. Graph.}} \bibinfo{volume}{31},
  \bibinfo{number}{4}, Article \bibinfo{articleno}{32} (\bibinfo{date}{July}
  \bibinfo{year}{2012}), \bibinfo{numpages}{12}~pages.
\newblock
\showISSN{0730-0301}
\urldef\tempurl%
\url{https://doi.org/10.1145/2185520.2185528}
\showDOI{\tempurl}


\bibitem[\protect\citeauthoryear{Ragan-Kelley, Barnes, Adams, Paris, Durand,
  and Amarasinghe}{Ragan-Kelley et~al\mbox{.}}{2013}]%
        {halide-pldi}
\bibfield{author}{\bibinfo{person}{Jonathan Ragan-Kelley},
  \bibinfo{person}{Connelly Barnes}, \bibinfo{person}{Andrew Adams},
  \bibinfo{person}{Sylvain Paris}, \bibinfo{person}{Fr\'{e}do Durand}, {and}
  \bibinfo{person}{Saman Amarasinghe}.} \bibinfo{year}{2013}\natexlab{}.
\newblock \showarticletitle{Halide: A Language and Compiler for Optimizing
  Parallelism, Locality, and Recomputation in Image Processing Pipelines}. In
  \bibinfo{booktitle}{\emph{Proceedings of the 34th ACM SIGPLAN Conference on
  Programming Language Design and Implementation}} (Seattle, Washington, USA)
  \emph{(\bibinfo{series}{PLDI '13})}. \bibinfo{publisher}{Association for
  Computing Machinery}, \bibinfo{address}{New York, NY, USA},
  \bibinfo{pages}{519–530}.
\newblock
\showISBNx{9781450320146}
\urldef\tempurl%
\url{https://doi.org/10.1145/2491956.2462176}
\showDOI{\tempurl}


\bibitem[\protect\citeauthoryear{Sanderson}{Sanderson}{2010}]%
        {armadillo}
\bibfield{author}{\bibinfo{person}{Conrad Sanderson}.}
  \bibinfo{year}{2010}\natexlab{}.
\newblock \bibinfo{booktitle}{\emph{{Armadillo: An Open Source C++ Linear
  Algebra Library for Fast Prototyping and Computationally Intensive
  Experiments}}}.
\newblock \bibinfo{type}{{T}echnical {R}eport}. \bibinfo{institution}{NICTA}.
\newblock


\bibitem[\protect\citeauthoryear{Senanayake, Hong, Wang, Wilson, Chou, Kamil,
  Amarasinghe, and Kjolstad}{Senanayake et~al\mbox{.}}{2020}]%
        {taco-ryan}
\bibfield{author}{\bibinfo{person}{Ryan Senanayake}, \bibinfo{person}{Changwan
  Hong}, \bibinfo{person}{Ziheng Wang}, \bibinfo{person}{Amalee Wilson},
  \bibinfo{person}{Stephen Chou}, \bibinfo{person}{Shoaib Kamil},
  \bibinfo{person}{Saman Amarasinghe}, {and} \bibinfo{person}{Fredrik
  Kjolstad}.} \bibinfo{year}{2020}\natexlab{}.
\newblock \showarticletitle{A Sparse Iteration Space Transformation Framework
  for Sparse Tensor Algebra}.
\newblock \bibinfo{journal}{\emph{Proc. ACM Program. Lang.}}
  \bibinfo{volume}{4}, \bibinfo{number}{OOPSLA}, Article
  \bibinfo{articleno}{158} (\bibinfo{date}{Nov.} \bibinfo{year}{2020}),
  \bibinfo{numpages}{30}~pages.
\newblock
\urldef\tempurl%
\url{https://doi.org/10.1145/3428226}
\showDOI{\tempurl}


\bibitem[\protect\citeauthoryear{Sorkine and Alexa}{Sorkine and Alexa}{2007}]%
        {arap}
\bibfield{author}{\bibinfo{person}{Olga Sorkine} {and} \bibinfo{person}{Marc
  Alexa}.} \bibinfo{year}{2007}\natexlab{}.
\newblock \showarticletitle{As-Rigid-As-Possible Surface Modeling}. In
  \bibinfo{booktitle}{\emph{Proceedings of EUROGRAPHICS/ACM SIGGRAPH Symposium
  on Geometry Processing}}. \bibinfo{pages}{109--116}.
\newblock


\bibitem[\protect\citeauthoryear{Tang, Liu, Zhao, Lin, Lin, Wang, and Han}{Tang
  et~al\mbox{.}}{2020a}]%
        {tang2020searching}
\bibfield{author}{\bibinfo{person}{Haotian* Tang}, \bibinfo{person}{Zhijian*
  Liu}, \bibinfo{person}{Shengyu Zhao}, \bibinfo{person}{Yujun Lin},
  \bibinfo{person}{Ji Lin}, \bibinfo{person}{Hanrui Wang}, {and}
  \bibinfo{person}{Song Han}.} \bibinfo{year}{2020}\natexlab{a}.
\newblock \showarticletitle{Searching Efficient 3D Architectures with Sparse
  Point-Voxel Convolution}. In \bibinfo{booktitle}{\emph{European Conference on
  Computer Vision}}.
\newblock


\bibitem[\protect\citeauthoryear{Tang, Schneider, Kamil, Panda, Li, and
  Panozzo}{Tang et~al\mbox{.}}{2020b}]%
        {eggs}
\bibfield{author}{\bibinfo{person}{Xuan Tang}, \bibinfo{person}{Teseo
  Schneider}, \bibinfo{person}{Shoaib Kamil}, \bibinfo{person}{Aurojit Panda},
  \bibinfo{person}{Jinyang Li}, {and} \bibinfo{person}{Daniele Panozzo}.}
  \bibinfo{year}{2020}\natexlab{b}.
\newblock \showarticletitle{EGGS: Sparsity-Specific Code Generation}.
\newblock \bibinfo{journal}{\emph{Computer Graphics Forum}}
  \bibinfo{volume}{39}, \bibinfo{number}{5} (\bibinfo{year}{2020}),
  \bibinfo{pages}{209--219}.
\newblock
\urldef\tempurl%
\url{https://doi.org/10.1111/cgf.14080}
\showDOI{\tempurl}
\showeprint{https://onlinelibrary.wiley.com/doi/pdf/10.1111/cgf.14080}


\bibitem[\protect\citeauthoryear{Van Der~Walt, Colbert, and Varoquaux}{Van
  Der~Walt et~al\mbox{.}}{2011}]%
        {numpy}
\bibfield{author}{\bibinfo{person}{Stefan Van Der~Walt},
  \bibinfo{person}{S~Chris Colbert}, {and} \bibinfo{person}{Gael Varoquaux}.}
  \bibinfo{year}{2011}\natexlab{}.
\newblock \showarticletitle{The NumPy array: a structure for efficient
  numerical computation}.
\newblock \bibinfo{journal}{\emph{Computing in Science \& Engineering}}
  \bibinfo{volume}{13}, \bibinfo{number}{2} (\bibinfo{year}{2011}),
  \bibinfo{pages}{22--30}.
\newblock


\bibitem[\protect\citeauthoryear{{V}an {Z}ee and {v}an~{d}e {G}eijn}{{V}an
  {Z}ee and {v}an~{d}e {G}eijn}{2015}]%
        {BLIS1}
\bibfield{author}{\bibinfo{person}{Field~G. {V}an {Z}ee} {and}
  \bibinfo{person}{Robert~A. {v}an~{d}e {G}eijn}.}
  \bibinfo{year}{2015}\natexlab{}.
\newblock \showarticletitle{{BLIS}: A Framework for Rapidly Instantiating
  {BLAS} Functionality}.
\newblock \bibinfo{journal}{\emph{ACM Trans. Math. Software}}
  \bibinfo{volume}{41}, \bibinfo{number}{3} (\bibinfo{date}{June}
  \bibinfo{year}{2015}), \bibinfo{pages}{14:1--14:33}.
\newblock
\urldef\tempurl%
\url{http://doi.acm.org/10.1145/2764454}
\showURL{%
\tempurl}


\bibitem[\protect\citeauthoryear{Venkat, Hall, and Strout}{Venkat
  et~al\mbox{.}}{2015}]%
        {chill}
\bibfield{author}{\bibinfo{person}{Anand Venkat}, \bibinfo{person}{Mary Hall},
  {and} \bibinfo{person}{Michelle Strout}.} \bibinfo{year}{2015}\natexlab{}.
\newblock \showarticletitle{Loop and Data Transformations for Sparse Matrix
  Code}. In \bibinfo{booktitle}{\emph{Proceedings of the 36th ACM SIGPLAN
  Conference on Programming Language Design and Implementation}} (Portland, OR,
  USA) \emph{(\bibinfo{series}{PLDI 2015})}. \bibinfo{pages}{521--532}.
\newblock


\bibitem[\protect\citeauthoryear{Venkat, Mohammadi, Rong, Barik, Park, Strout,
  and Hall.}{Venkat et~al\mbox{.}}{2016}]%
        {venkat2016}
\bibfield{author}{\bibinfo{person}{Anand Venkat}, \bibinfo{person}{Mahdi~Soltan
  Mohammadi}, \bibinfo{person}{Hongbo Rong}, \bibinfo{person}{Rajkishore
  Barik}, \bibinfo{person}{Jongsoo Park}, \bibinfo{person}{Michelle~Mills
  Strout}, {and} \bibinfo{person}{Mary Hall.}} \bibinfo{year}{2016}\natexlab{}.
\newblock \showarticletitle{Automating Wavefront Parallelization for Sparse
  Matrix Computations}. In \bibinfo{booktitle}{\emph{In Supercomputing (SC)}}.
\newblock


\bibitem[\protect\citeauthoryear{Virtanen, Gommers, Oliphant, Haberland, Reddy,
  Cournapeau, Burovski, Peterson, Weckesser, Bright, {van der Walt}, Brett,
  Wilson, Millman, Mayorov, Nelson, Jones, Kern, Larson, Carey, Polat, Feng,
  Moore, {VanderPlas}, Laxalde, Perktold, Cimrman, Henriksen, Quintero, Harris,
  Archibald, Ribeiro, Pedregosa, {van Mulbregt}, and {SciPy 1.0
  Contributors}}{Virtanen et~al\mbox{.}}{2020}]%
        {scipy}
\bibfield{author}{\bibinfo{person}{Pauli Virtanen}, \bibinfo{person}{Ralf
  Gommers}, \bibinfo{person}{Travis~E. Oliphant}, \bibinfo{person}{Matt
  Haberland}, \bibinfo{person}{Tyler Reddy}, \bibinfo{person}{David
  Cournapeau}, \bibinfo{person}{Evgeni Burovski}, \bibinfo{person}{Pearu
  Peterson}, \bibinfo{person}{Warren Weckesser}, \bibinfo{person}{Jonathan
  Bright}, \bibinfo{person}{St{\'e}fan~J. {van der Walt}},
  \bibinfo{person}{Matthew Brett}, \bibinfo{person}{Joshua Wilson},
  \bibinfo{person}{K.~Jarrod Millman}, \bibinfo{person}{Nikolay Mayorov},
  \bibinfo{person}{Andrew R.~J. Nelson}, \bibinfo{person}{Eric Jones},
  \bibinfo{person}{Robert Kern}, \bibinfo{person}{Eric Larson},
  \bibinfo{person}{C~J Carey}, \bibinfo{person}{{\.I}lhan Polat},
  \bibinfo{person}{Yu Feng}, \bibinfo{person}{Eric~W. Moore},
  \bibinfo{person}{Jake {VanderPlas}}, \bibinfo{person}{Denis Laxalde},
  \bibinfo{person}{Josef Perktold}, \bibinfo{person}{Robert Cimrman},
  \bibinfo{person}{Ian Henriksen}, \bibinfo{person}{E.~A. Quintero},
  \bibinfo{person}{Charles~R. Harris}, \bibinfo{person}{Anne~M. Archibald},
  \bibinfo{person}{Ant{\^o}nio~H. Ribeiro}, \bibinfo{person}{Fabian Pedregosa},
  \bibinfo{person}{Paul {van Mulbregt}}, {and} \bibinfo{person}{{SciPy 1.0
  Contributors}}.} \bibinfo{year}{2020}\natexlab{}.
\newblock \showarticletitle{{{SciPy} 1.0: Fundamental Algorithms for Scientific
  Computing in Python}}.
\newblock \bibinfo{journal}{\emph{Nature Methods}}  \bibinfo{volume}{17}
  (\bibinfo{year}{2020}), \bibinfo{pages}{261--272}.
\newblock
\urldef\tempurl%
\url{https://doi.org/10.1038/s41592-019-0686-2}
\showDOI{\tempurl}


\bibitem[\protect\citeauthoryear{Vuduc, Demmel, and Yelick}{Vuduc
  et~al\mbox{.}}{2005}]%
        {oski}
\bibfield{author}{\bibinfo{person}{Richard Vuduc}, \bibinfo{person}{James~W.
  Demmel}, {and} \bibinfo{person}{Katherine~A. Yelick}.}
  \bibinfo{year}{2005}\natexlab{}.
\newblock \showarticletitle{{OSKI: A library of automatically tuned sparse
  matrix kernels}}.
\newblock \bibinfo{journal}{\emph{Journal of Physics: Conference Series}}
  \bibinfo{volume}{16}, \bibinfo{number}{1} (\bibinfo{year}{2005}),
  \bibinfo{pages}{521+}.
\newblock
\showISSN{1742-6596}


\bibitem[\protect\citeauthoryear{Walter and Koch}{Walter and Koch}{2007}]%
        {ublas}
\bibfield{author}{\bibinfo{person}{Joerg Walter} {and} \bibinfo{person}{Mathias
  Koch}.} \bibinfo{year}{2007}\natexlab{}.
\newblock \bibinfo{title}{{uBLAS}}.
\newblock
\newblock
\urldef\tempurl%
\url{http://www.boost.org/libs/numeric/ublas/doc/index.htm}
\showURL{%
\tempurl}


\bibitem[\protect\citeauthoryear{Whaley and Dongarra}{Whaley and
  Dongarra}{1998}]%
        {atlas}
\bibfield{author}{\bibinfo{person}{R.~Clint Whaley} {and} \bibinfo{person}{Jack
  Dongarra}.} \bibinfo{year}{1998}\natexlab{}.
\newblock \showarticletitle{Automatically Tuned Linear Algebra Software}. In
  \bibinfo{booktitle}{\emph{SuperComputing 1998: High Performance Networking
  and Computing}}.
\newblock


\bibitem[\protect\citeauthoryear{Wolff, Herholz, Ziegler, Link, Brügel, and
  Sorkine-Hornung}{Wolff et~al\mbox{.}}{2021}]%
        {Wolff:2021}
\bibfield{author}{\bibinfo{person}{Katja Wolff}, \bibinfo{person}{Philipp
  Herholz}, \bibinfo{person}{Verena Ziegler}, \bibinfo{person}{Frauke Link},
  \bibinfo{person}{Nico Brügel}, {and} \bibinfo{person}{Olga
  Sorkine-Hornung}.} \bibinfo{year}{2021}\natexlab{}.
\newblock \bibinfo{title}{3D Custom Fit Garment Design with Body Movement}.
\newblock
\newblock
\showeprint[arxiv]{2102.05462}~[cs.GR]


\end{thebibliography}

\appendix
\section{Algebraic hashing} \label{app:algHash}%
\begin{algorithm}[h!]%
    \AlgName{algebraicHash(SymbolicExpression x)}\\
\KwOut{\textrm{Algebraic hash value for expression x.}}
\vspace{4pt}

\uIf(\tcp*[f]{return hash if it has been computed.}){hashIsValid}{\Return h}
\Else{

\uIf{Op(x) == Variable}{h = Hash(VariableIndex(x))}
 \uElseIf{Op(x) == Constant}{
 \uIf{isSmallInteger(Value(x))}{h = Value(x)}
\Else{h = Hash(Value(x))}
}
\uElseIf{Op(x) == Multiplication}
 {h = 1\\
  \ForEach{c in Childs(x)}
 {h *= algebraicHash(c)}
 }
 \uElseIf{Op(x) == Addition}
 {h = 0\\
  \ForEach{c in Childs(x)}
 {h += algebraicHash(c)}
 }
 
 \Else{
 \ForEach{c in Childs(x)}
 {h = Hash(algebraicHash(c), h)}}

 h = Hash(Hash(Op(x)), h)\\
   \Return h
 }
    \caption{Algebraic Hashing}%
    \label{alg:algebraichashing}%
\end{algorithm}%
\noindent Algorithm \ref{alg:algebraichashing} presents pseudo code for our algebraic hashing procedure descibed in \secref{sec:algHash}. If two expressions have the same algebraic hash value, they will evaluate to the same value; however, there are equivalent expressions that can not be detected like $\sin(x + \pi / 2)$ and $\cos(x)$.

\section{Leaf harvesting example}\label{app:lhe}
In the illustrated example (\figref{fig:treeTemplate}), leaf harvesting would result in the following array of data which contains the specific variable or constant stored in the $6$ leaves of each group member.
\begin{center}
    \begin{tabular}{lllllll}
               & $l_0$ & $l_1$ & $l_2$ & $l_3$ & $l_4$ & $l_5$ \\\toprule
        tree 1 & b     & b     & 2     & a     & 3.1   & a     \\
        tree 2 & c     & c     & 2     & d     & 2.7   & d     \\\hline
               & $x_0$ & $x_0$ & 2     & $x_1$ & $c_0$ & $x_1$
    \end{tabular}
    \hspace{2em}
    \begin{tabular}{lll}
        $x_0$       & $x_1$       & $c_0$         \\\toprule
        b           & a           & 3.1           \\
        c           & d           & 2.7           \\\hline
        \phantom{b} & \phantom{a} & \phantom{3.1}
    \end{tabular}
\end{center}
From this data, we can see that leaves $l_0$, and $l_1$ always contain the same variable. The same is true for $l_3$ and $l_5$. Leaf $l_3$ always contains the \emph{uniform} constant $2$ which can be hard-coded and leaf $l_4$ contains a \emph{varying} constant that has to be read from memory. This analysis will reduce the use of memory bandwidth since we can avoid redundant loads or even avoid them completely (\secref{sec:implementation}).

\section{Automatic differentiation}
\label{sec:appendixAutoDiff}
\begin{algorithm}[h!]
    \SetKw{kwNot}{not}
\AlgName{differentiate(Symbolic x)}\\
\KwOut{\textrm{Gradient of the expression x.}}
\vspace{4pt}

grad = (0, ..., 0)\\
stack $\leftarrow$ (x, 1)\\

\While{\kwNot isEmpty(stack)} {
	y = pop(stack)\\

	\Switch{op(y[0])} {
		\Case{Add} {
			stack $\leftarrow$ (left(y[0]), y[1])\\
			stack $\leftarrow$ (right(y[0]), y[1])
		}
		\Case{Sub} {
			stack $\leftarrow$ (left(y[0]), y[1])\\
			stack $\leftarrow$ (right(y[0]), -y[1])
		}
		\Case{Mul} {
			stack $\leftarrow$ (left(y[0]), y[1] * right(y[0]))\\
			stack $\leftarrow$ (right(y[0]), y[1] * left(y[0]))
		}
		\Case{Div} {
			stack $\leftarrow$ (left(y[0]), y[1] / right(y[0]))\\
			stack $\leftarrow$ (right(y[0]), -left(y[0]) / (right(y[0]) * right(y[0])) * y[1])
		}
		\Case{VAR} {
			grad[varId(y[0])] += y[1]
		}
	}
}

\Return grad

    \caption{Differentiation of a symbolic expression.}
    \label{alg:autodiff}
\end{algorithm}

In Algorithm \ref{alg:autodiff} we outline a simple version of our reverse-mode differentiation implementation following the concise treatment of Baydin et al.~\shortcite{baydin2018automatic}. \revv{The listing illustrates that it is quite simple to add symbolic differentiation capabilities based on our \texttt{Symbolic} type. Moreover, adding support for additional operations is trivial and amounts to adding a \texttt{case} statement that defines the derivative of an operation with respect to the operands. The algorithm traverses the derivatives from the expression tree root down to it's leaves motivating the name reverse-mode differentiation.} The functions \texttt{left} and \texttt{right} access the first and second child of an expression; the operator $\leftarrow$ pushes elements onto the stack. 
\begin{table*}
    \begin{center}
\begin{tabular}{lllrrrrrr}
figure & experiment & size & execute code & analyze expressions & generate & \revv{peak} Mem. &compile (gcc) & binary  \\\toprule
\ref{fig:sparseSpeedup} & numeric\_0 & 100k & 1m 21s 458ms & 2m 54s 911ms & 1s 472ms & 7.7GB& 1s 149ms & 36KB\\
\ref{fig:sparseSpeedup} &  numeric\_1 & 100k & 2m 3s 939ms & 4m 35s 377ms & 2s 435ms &11.7GB& 1s 122ms& 36KB\\
\ref{fig:sparseSpeedup} &  numeric\_2 & 100k & 1m 54s 64ms & 3m 30s 903ms & 1s 773ms &10.1GB& 1s 88ms& 36KB\\
\ref{fig:sparseSpeedup} &  numeric\_0 & 500k & 7m 41s 301ms & 14m 55s 165ms & 8s 564ms &38.7GB& 1s 97ms& 36KB\\
\ref{fig:sparseSpeedup} &  numeric\_1 & 500k & 11m 40s 599ms & 22m 20s 401ms & 11s 797ms &58.1GB& 1s 58ms& 36KB\\
\ref{fig:sparseSpeedup} &  numeric\_2 & 500k & 10m 56s 974ms & 17m 14s 999ms & 9s 547ms &50.7GB& 1s 75ms& 36KB\\
\ref{fig:sparseSpeedup} &  numeric\_0 & 1M & 15m 12s 281ms & 30m 5s 612ms & 17s 283ms &77.4GB& 1s 43ms& 36KB\\
\ref{fig:sparseSpeedup} &  numeric\_1 & 1M & 26m 5s 690ms & 47m 6s 461ms & 26s 563ms &116.2GB& 1s 52ms & 36KB\\
\ref{fig:sparseSpeedup} &  numeric\_2 & 1M & 21m 58s 634ms & 34m 25s 969ms & 20s 128ms &99.6GB& 1s 49ms & 36KB\\
\ref{fig:cotan}  &  cotan & 1M & 1m 39s 345ms & 11m 27s 881ms & 8s 300ms &40.9GB& 1s 544ms & 57KB\\
\ref{fig:cotan}  &  cotan & 500k & 48s 568ms & 5m 24s 992ms & 4s 131ms &19.9GB& 1s 490ms & 57KB\\
\ref{fig:cotan}  &  cotan & 200k & 20s 607ms & 2m 17s 935ms & 1s 930ms &8.2GB& 1s 473ms & 57KB\\
\ref{fig:dualLaplace} &  dual Laplace & 100k & 5m 1s 537ms & 42m 14s 675ms & 24s 175ms &136.0GB& 7s 421ms  & 343KB\\
\ref{fig:dualLaplace} &  dual Laplace & 20k & 47s 852ms & 7m 15s 740ms & 4s 356ms &29.6GB& 7s 473ms & 327KB\\
\ref{fig:dualLaplace} &  dual Laplace & 7k & 16s 49ms & 2m 23s 564ms & 1s 415ms &10.0GB& 7s 503ms & 258KB\\
\ref{fig:clothAD} &  cloth Hessian AD & 100k & 19m 46s 671ms & 6m 37s 590ms & 83ms &122.9GB& 1s 509ms  & 57KB\\
\ref{fig:clothAD} &  cloth Hessian AD & 50k & 9m 6s 206ms & 2m 25s 926ms & 37ms &56.6GB& 1s 603ms & 57KB\\
\ref{fig:clothAD} &  cloth Hessian AD & 20k & 4m 2s 526ms & 51s 499ms & 17ms &25.0GB& 1s 516ms & 57KB\\
\ref{fig:clothAD} &  cloth Hessian direct & 100k & 2m 44s 437ms & 3m 5s 543ms & 143ms &18.4GB& 1s 515ms  & 61KB\\
\ref{fig:clothAD} &  cloth Hessian direct & 50k & 1m 31s 336ms & 1m 12s 900ms & 49ms &8.5GB& 1s 397ms & 58KB\\
\ref{fig:clothAD} &  cloth Hessian direct & 20k & 33s 512ms & 24s 217ms & 25ms &3.7GB& 1s 587ms  & 58KB\\
\ref{fig:adCompare} & symmetric Dirichlet & 1M & 7m 0s 371ms & 8m 26s 669ms & 556ms &135.2GB& 1s 272ms  & 61KB\\
\ref{fig:arapRHS} &  ARAP rhs & 300k & 5m 12s 580ms & 6m 16s 664ms & 506ms &34.3GB& 3s 127ms  & 248KB\\
\ref{fig:clothTimes} & cloth full system & 100k & 4m 36s 584ms & 6m 10s 303ms & 6m 10s 303ms &37.0GB& 4s 521ms  & 207KB\\
\ref{fig:clothTimes} & cloth full system& 50k & 2m 7s 959ms & 3m 32s 458ms & 3m 32s 458ms &17.0GB& 4s 543ms  & 135KB\\
\ref{fig:clothTimes} & cloth full system & 20k & 55s 56ms & 57s 404ms & 57s 404ms &7.5GB& 4s 521ms & 131KB\\\hline
\end{tabular}
\end{center}

    \caption{We break down system performance for different stages of our method as well as the use of resources for all our examples.}
    \label{tab:timings}
\end{table*}
\section{Expression simplification}\label{app:simplifcationRules}
We implemented the following set of transformations, and we introduce each of them with a typical example.

\paragraph{Factorization}
For every sum of products we find the largest set of factors that is contained in as many factors as possible. This enables the following algebraic transformation:
\begin{align*}
    x y z + x w y + v x y + ux & \rightarrow x y (z + w + v) + u x  \\
                               & \rightarrow x (y (z + w + v) + u).
\end{align*}
The algorithm tries to factor out the largest expression $xy$ first even though it is not contained in all summands. Recursively calling the factorization routine allows us to further simplify the expression by factoring out $x$ as well.

\paragraph{Reducing fractions}
We analyze each subtree that contains only multiplications and divisions. By first expanding it into a pure product we can perform the following type of simplification:
\begin{align*}
    \frac{(x + y) z^2}{(x + y)^2 w} \frac {w^2}{z} & \rightarrow (x + y) z^2 \frac{1}{(x + y)^2}  \frac{1}{w} w^2 \frac{1}{z}
    \rightarrow  \frac{z w}{(x+y)}.
\end{align*}
In this step, we also find reciprocals of square roots, which allows us to compute them using fast implementations like Cuda's \texttt{rsqrt} if available.

\paragraph{Summand elimination}
By analyzing all subtrees of additions in a similar way, we can perform the transformation
\begin{align*}
    (x + x + y - 2 (y + x)) \rightarrow 2 x + y + (-2 y) + (-2 x) \rightarrow -y.
\end{align*}
All products of constants with sums are expanded to group multiples of expressions. Negative sums are treated as the product of a sum with the constant $-1$.

\paragraph{Square root elimination}
Square roots commonly appear when normalizing vectors or computing angles between them.
We consolidate products and divisions of square roots in order to minimize computational effort. We can also eliminate square roots of squares that would appear in a naive implementation of the squared norm
\begin{align*}
    \left\|\begin{pmatrix} x \\ y \end{pmatrix} \right\|^2 = \sqrt{x^2 + y^2}\sqrt{x^2 + y^2} \rightarrow \sqrt{(x^2 + y^2)(x^2 + y^2)} \rightarrow x^2 + y^2.
\end{align*}

\paragraph{Constant expressions}
Arithmetic hashing allows us to predict if a complex expression will evaluate to a constant value. This gives us a powerful tool to simplify expressions that cannot be explicitly transformed into a simpler version but still evaluate to a constant. The basic transformation
\begin{align*}
    \frac {(x - y)^2 + xy} {x^2 - xy + y^2} \rightarrow 1
\end{align*}
is only possible because the value hash of the expression evaluates to one.

\section{Data structures} \label{app:type}
Our implementation is build around the \texttt{Symbolic} type that is replacing floating point values. A basic, unoptimized definition of this type is presented in the following listing.
\vspace{4pt}
\begin{minted}{cpp}
class Symbolic {

    struct SData {
        int opCode;
        vector<Symbolic> childs;
        int variable_id;
        double constant;

        size_t structureHash();
        size_t algebraicHash();
        unsigned complexity();
    };

    shared_ptr<SData> data;

public:
    Symbolic(int opCode, vector<Symbolic>& operands);
    Symbolic(double constant);
    Symbolic(int variableType, int variableId);
    Symbolic operator+(const Symbolic& b);
    Symbolic operator+=(const Symbolic& b);
    ...
};

Symbolic operator+(const double f, Symbolic& b);
Symbolic operator+(Symbolic& b, const double f);
Symbolic operator+=(Symbolic& b, const double f);
Symbolic sqrt(Symbolic& s);  
...
\end{minted}
\vspace{12pt}
The \texttt{Symbolic} type itself only stores a shared pointer to the actual data which is implemented as a nested type. The class \texttt{Symbolic} is responsible for overloading the required arithmetic operators while library functions and arithmetic operations involving floating point values are overloaded externally. Using the shared pointers makes memory management convenient and allows to save on memory as trees representing a local variable are stored only once even though the variable might be involved in different operations.

\section{Code optimization timings}
\label{sec:codeTimings}
In Table \ref{tab:timings} we show a breakdown of the time and resources our system needed in order to generate the final, optimized program for all examples shown in the paper. The timings do not depend on the target architecture. The \emph{execution} phase runs the original program with symbolic types.
The \emph{analysis} includes expression decomposition and grouping, leaf harvesting, and expression optimization. Code \emph{generation} writes the code file and constructs all position indices for indirect memory access. \emph{Compilation} creates the final program based on the generated code (we use gcc in this case). Most time is spent analyzing the expressions and their dependencies since all subexpressions above the complexity threshold have to be considered, indexed, and eventually decomposed. Compilation times and the final binary size are largely independent of the problem size which is expected since the expression will decompose into very similar sets of groups; they will just contain more instances. A crucial limitation of our method is the high memory demand for storing all expression trees, however, we show examples with up to a million vertices which is, depending on the context, relatively large. Our code is in large parts not optimized and would benefit from parallelization and advanced memory optimization.

\end{document}